\documentclass[journal]{IEEEtran}
\usepackage{array}
\usepackage{textcomp}
\usepackage{stfloats}
\usepackage{amsmath,amssymb,amsfonts}
\usepackage{textcomp}
\usepackage{xcolor}
\usepackage{subfigure}
\usepackage[noend]{algpseudocode}
\usepackage{algorithm}
\usepackage{algorithmicx}
\usepackage{booktabs}
\usepackage{makecell}
\usepackage{multirow}
\usepackage{threeparttable}
\usepackage{framed}
\usepackage{graphicx}  
\usepackage{epstopdf}
\usepackage{url}
\usepackage{hyperref}

\begin{document}

\title{Understanding the Decentralization of DPoS: Perspectives From Data-Driven Analysis on EOSIO}

\author{Jieli~Liu, Weilin Zheng, Dingyuan Lu, Jiajing~Wu,~\IEEEmembership{Senior Member,~IEEE,}
	and~Zibin~Zheng,~\IEEEmembership{Senior Member,~IEEE}
	\thanks{Manuscript received January xx, 2022. The research is supported by xxx). (\textit{Corresponding Author: Jiajing Wu})}
	\thanks{J. Liu and Z. Zheng are with the School of Software Engineering, Sun Yat-sen University, Zhuhai 519082, China. W. Zheng, D. Lu, and J. Wu are with the School of Computer Science and Engineering, Sun Yat-sen University, Guangzhou 510006, China. (Email: liujli7@mail2.sysu.edu.cn, wujiajing@mail.sysu.edu.cn)}
}

\markboth{Journal of \LaTeX\ Class Files,~Vol.~14, No.~8, August~2021}%
{Shell \MakeLowercase{\textit{et al.}}: A Sample Article Using IEEEtran.cls for IEEE Journals}

\IEEEpubid{0000--0000/00\$00.00~\copyright~2021 IEEE}

\maketitle

\begin{abstract}
Recently, many Delegated Proof-of-Stake (DPoS)-based blockchains have been widely used in decentralized applications, such as EOSIO, Tron, and Binance Smart Chain. Compared with traditional PoW-based blockchain systems, these systems achieve a higher transaction throughput and are well adapted to large-scale scenes in daily applications. Decentralization is a key element in blockchain networks. However, little is known about the evolution of decentralization in DPoS-based blockchain networks. In this paper, we conduct a systematic analysis on the decentralization of DPoS with data from up to 135,000,000 blocks in EOSIO, the first successful DPoS-based blockchain system. We characterize the decentralization evolution of the two phases in DPoS, namely block producer election and block production. Moreover, we study the voters with similar voting behaviors and propose methods to discover abnormal mutual voting behaviors in EOSIO. The analytical results show that our methods can effectively capture the decentralization evolution and abnormal voting phenomena in the system, which also have reference significance for other DPoS-based blockchains.  

\end{abstract}


\section{Introduction}\label{Introduction}

\IEEEPARstart{N}{owadays}, blockchain has aroused great attention among people and has been widely applied in many fields such as finance, healthcare, and Internet of Things (IoTs). Technically, blockchain \cite{zheng2017overview} is an append-only distributed ledger database combined with hybrid techniques like peer-to-peer networks, cryptography, and consensus mechanisms. Different from traditional centralized systems, blockchain provides users a decentralized environment that can avoid the single point failure of a system. Thus decentralization is a core element in blockchain networks.

The most famous blockchain system is Bitcoin \cite{nakamoto2019bitcoin} and Ethereum \cite{wood2014ethereum}, which is based on the Proof-of-Work (PoW) consensus protocol~\cite{finney2004rpow}. However, with the wide application of blockchain, both Bitcoin and Ethereum suffer from the low scalability problem and can not meet the growing application requirement. The Delegated Proof-of-Stake (DPoS) consensus protocol is an efficient and flexible blockchain consensus mechanism famous for its high scalability in block production. Unlike traditional PoW-based consensus protocols, DPoS concentrates the block production process in the hands of a small set of block producers (also called super nodes) elected by the entire network, and thus improves the transaction throughput. The original DPoS was proposed in 2014 \cite{larimer2014delegated}, and it has given rise to many variant versions adopted in a series of successful blockchains like EOSIO \cite{io2017eos}, Tron \cite{Tronwhitepaper} and Binance Smart Chain \cite{bsc2020}. 
The framework of all these DPoS consensus protocols can be summarized into two phases: 
\begin{itemize}
	\item[1)] Block producer election: This phase dynamically elects a limited set of block producers to produce blocks. 
	The voting weights are different across the stakes of voters. A fixed number of top candidates receiving the highest voting weight can become block producers.
	\item[2)] Block production: The elected block producers begin a round of block production orderly. Each elected block producer has a fixed time slot to produce blocks. If a block is not produced at the scheduled time, this block will be skipped and will not affect the subsequent block production. Finally, the block production rewards are distributed to these elected block producers and a new round of block producer election and block production will be started.
\end{itemize}

\IEEEpubidadjcol

Out of the importance of blockchain security, recently many efforts have been devoted to analyzing the decentralization in different blockchain systems. Intuitively, if a blockchain system is not decentralized enough, it can be easily controlled by a small minority of parties. In this case, the blockchain system is fragile and cannot ensure the facticity of the ledger records when facing threats like 51\% attack \cite{ye2018analysis} and selfish mining \cite{eyal2014majority}. Existing research \cite{10.1007/978-3-662-58387-6_24,9160462} studied the decentralization of Bitcoin and Ethereum in terms of mining power, bandwidth, etc.
For DPoS-based blockchains, some studies \cite{xu2018eos,rebellosecurity} have theoretically analyzed the decentralization from the protocols and pointed out several clear vulnerabilities. Some other studies \cite{10.1145/3318041.3355463,10.1007/978-3-030-59638-5_2} measured the decentralization of blockchain systems with the entropy metric. This metric is useful in comparing the decentralization among different protocols according to the block production records of blockchain systems. However, there is no study quantifying the decentralization evolution in DPoS-based blockchains. And an in-depth study on the behaviors of voters and block producers can help us timely capture the abnormal phenomena such as voting manipulation in a DPoS-based blockchain system. Recent reports \cite{xu2018eos,Huobidataleak} have pointed out that it is possible to achieve voting collusion in DPoS-based blockchains. Yet there is still a lack of study to capture the abnormal voting behaviors in DPoS-based blockchain systems.

\begin{figure}[t]
	\centerline{\includegraphics[scale=0.34]{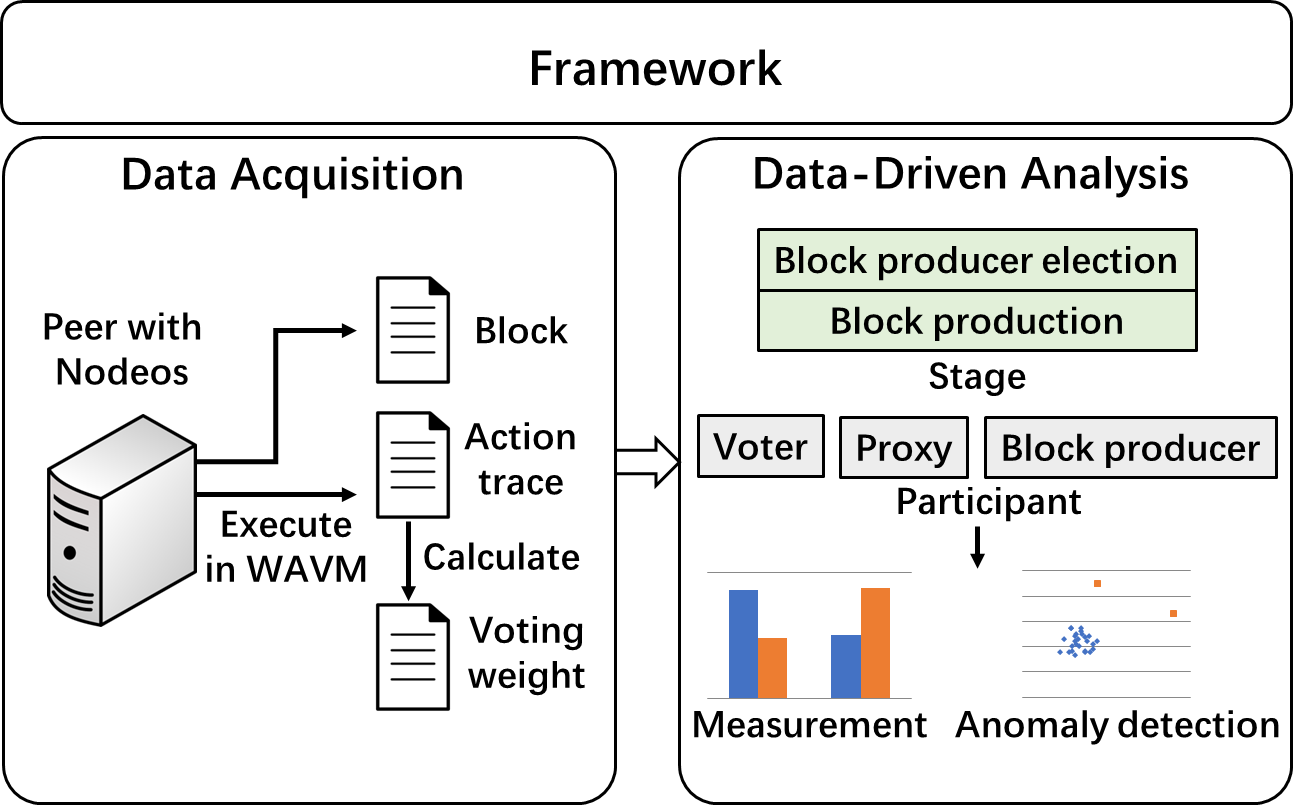}}
	\caption{The analysis framework.}
	\label{fig_framework}
\end{figure}

To fill this gap, we conduct a data-driven analysis with the blockchain data from EOSIO. EOSIO \cite{io2017eos} is a typical and the first successful DPoS-based blockchain system. Up to now, the number of transactions in EOSIO has exceeded 5.0 billion according to the statistic of \textit{eosflare.io}~\cite{eosflare}. Our analysis framework is shown in Fig. \ref{fig_framework}. Firstly, we collect the block data as well as the action trace data, and calculate the voting weight data according to the action trace data. Then in the analysis, we conduct a decentralization evolution measurement by characterizing all voters, voting proxies, and block producers participating in the DPoS consensus process. Following the analysis, we design network-based methods to uncover the abnormal voting behaviors in the system. In particular, we investigate the abnormal voting gangs with similar voting behaviors and mutual voting behaviors. The analytical results show that our study can reveal the decentralization evolution and effectively capture some abnormal voting activities in a DPoS-based blockchain.

In summary, we make the following contributions in this paper:
\begin{itemize}
	\item 
	To the best of our knowledge, our work is the first data-driven study on the decentralization evolution of DPoS. Based on the data from a typical DPoS-based system named EOSIO, we quantify its decentralization and capture the abnormal voting behaviors in the system.
	\item We report some voting gang anomalies with our methods, including gangs with similar voting behaviors and mutual voting behaviors. Moreover, we also analyze the phenomena behind the detected results.
	\item We find that the EOSIO proxies account for an increasingly large proportion of the total voting power, and the set of block producers changes less and less. Besides, the overall decentralization in EOSIO is also affected by these factors. Our findings and discussion can provide a reference for other DPoS-based blockchains.
\end{itemize}

The remaining sections of this paper are organized as follows. Section \ref{Background} and Section \ref{DC} first provide the background and detail the data collection. Section \ref{DMS} and Section \ref{VMD} next introduce the decentralization measurement study and analysis on the abnormal voting phenomena. Section \ref{implications} discusses some implications from our findings for DPoS-based blockchains and Section \ref{RW} provides the related work. Finally, we conclude the paper and discuss the future work in Section \ref{Conclusion}.

\section{Background}\label{Background}

In this section, we provide the background required to understand the operation mechanism of EOSIO and the DPoS consensus in EOSIO.

\subsection{Background of EOSIO}
EOSIO is a blockchain system built for providing an operating environment for a large scale of commercial DApps, being regarded as the foundation of blockchain 3.0 \cite{xu2018eos}. Different from the most famous blockchain platforms like Bitcoin and Ethereum, EOSIO adopts DPoS as its consensus mechanism, which allows it to achieve a far higher transaction throughput. Recent years have seen a growth in the popularity of EOSIO in DApp transactions since its high performance and a waiver of transaction fees.

Unlike Ethereum, the identity of an account in EOSIO is a human-defined string with up to 12 characters in length. Each account is created by an existing account via the \textit{newaccount} interface and can deploy a contract on itself via the \textit{setcode} interface of the system account \textit{eosio}. The main currency circulating in EOSIO is EOS tokens, and resources of smart contracts such as CPU and RAM can be obtained via EOS token mortgage. A transaction in EOSIO contains multiple actions, and each action corresponds to an invocation of a contract. According to the initiator in contract invocation, these actions can be divided into calling actions and inline actions, for which are called by users and triggered by contracts respectively~\cite{xu2018eos}. Besides, EOSIO provides a flexible role-based permission management allowing users to delegate their permissions to achieve a high-level control on other users. 

\subsection{DPoS and DPoS in EOSIO}
The DPoS consensus has become increasingly popular and widely adopted in several successful blockchain systems such as EOSIO, Tron and Binance Smart Chain (BSC). Instead of solving the PoW puzzles, DPoS decides its block producers according to the votes of the entire stakeholders, thus achieving high scalability in block production. The DPoS consensus consists of the block producer election phase and the block production phase, and the details of these in EOSIO are provided as follows: 

\subsubsection{Phase 1: Block Producer Election}
This phase elects 21 block producers for block production. To become a block producer candidate, anyone in EOSIO who possesses enough hardware resources and the full ledger data can register via the \textit{regproducer} interface of \textit{eosio}. And the top 21 candidates with the highest voting weight can become block producers and obtain block production rewards.

Anyone in EOSIO can vote for the block producer candidates in two ways. The first way is voting directly through setting the \textit{producers} parameter of the \textit{voteproducer} interface as the list of at most 30 selected candidates. The second way is voting through a proxy by setting the \textit{proxy} parameter of the \textit{voteproducer} interface as the chosen proxy, and then the proxy can vote for block producer candidates on behalf of all proxied users. A user can register to become a voting proxy by setting the \textit{isproxy} parameter of the \textit{regproxy} interface as 1, otherwise, a user can cancel the registration by setting the \textit{isproxy} parameter as 0.

The voting weight of a voter is related to the voting time and the staked tokens for CPU and bandwidth, which can be calculated as:
\begin{equation}
weight = 10000 * stake * 2 ^{index},
\end{equation}
where $stake$ is equal to the amount of delegated bandwidth and computation, and $index$ can grow incrementally every week if voters update their voting, calculated as:
$
index = \frac{1}{52} * \lfloor\frac{t_{vote} - t_{init}}{7 * t_{day}}\rfloor.
$
In this formula, $t_{vote}$ represents the Unix timestamp when a user performs the \textit{voteproducer} operation, $t_{init}$ represents the Unix timestamp of Jan. 1, 2000, and $t_{day}$ denotes the number of seconds per day. As we can see, with the same amount of stakes, the voting weight of recent votings is higher than that of earlier votings. This design can encourage voters to keep their votes updated. Besides, once a user updates the number of stakes via \textit{delegatebw} (increase mortgage) or \textit{undelegatebw} (reduce mortgage), the voting weight will be automatically updated. For a voting proxy, its voting weight will be the sum of voting weight owned by all proxied voters. If a voter votes for multiple candidates, each chosen candidate can receive an equal voting weight according to the stakes and voting time of the voter. 

\subsubsection{Phase 2: Block Production}
In this phase, the elected 21 block producers from the block producer election begin a round of block production on behalf of the stakeholders in EOSIO. They validate the transactions, construct the valid transactions into blocks and then produce blocks orderly. Each elected block producer has a fixed 3-second to produce 6 blocks. Therefore, a round of block production lasts 63 seconds. If a block is not produced at the scheduled time because of network delay or other reasons, this block will be skipped and will not affect the subsequent block production. The new block confirmation process is also conducted among the 21 block producers. Once a new block is confirmed by more than 2/3 of the block producers (i.e. at least $21\times 2/3 +1 = 15$ block producers) through signed messages, this block will be appended to the blockchain. With DPoS, EOSIO can generate a new block every 0.5 seconds averagely with a throughput of up to 8,000 TPS \cite{10.1145/3392155}, which achieves higher performance than many other blockchain systems.

\section{Data Collection}\label{DC}
We conduct our experiments on up to 135,000,000 blocks in EOSIO, which cover the transaction data from Jun. 8, 2018 to Aug. 5, 2020. To measure the decentralization of EOSIO, we collect three types of blockchain data, including block header, transaction actions, and the voting weight data by running an EOSIO full node and replaying all transactions.

\begin{itemize} 
	\item \textbf{Block header:} The block header in each block provides a summary for the entire block, which includes information such as the block producer, the block timestamp, the transaction Merkle root, etc. We obtain the block header data by starting a core service daemon of EOSIO named \textit{Nodeos} \cite{Nodeos} to synchronize data on the mainnet. 
	
	\item \textbf{Transaction actions:} A transaction in EOSIO contains a set of actions, and each action is an invocation of a smart contract. However, only the calling actions are recorded in the blockchain, which are operations performed by users. While the inline actions, which are triggered by calling actions, are not recorded in the on-chain data and can be obtained only by replaying all transactions. To obtain the full transaction action data, we make use of the action trace data, which are the detailed run-time data of smart contract invocation generated by the Web Assembly Virtual Machine (WAVM). With \textit{history\_file\_plugin} \cite{zheng2020xblock}, we collect all action traces in JSON format. Then we extract the related actions such as \textit{delegatebw}, \textit{undelegatebw}, \textit{regproxy}, \textit{voteproducer}, etc from the action traces.
	
	\item \textbf{Voting weight data:} The voting weight received by each block producer candidate is in constant change since it is affected by the change of voters and their stakes. Thus we calculate the voting weight data of each candidate by traversing the transaction actions and record the data in periodically. 
	
\end{itemize}

\section{Decentralization Measurement Study}\label{DMS}
In this section, we present a decentralization measurement study in EOSIO considering the processes of block producer election and block production. We try to characterize all voters, proxies, and block producers participating in the DPOS consensus process. Based on the analysis results, we reveal the evolution of decentralization and  discuss some findings in the EOSIO network. 

\subsection{Block Producer Election in EOSIO} \label{BPEE}
\subsubsection{Overview of Block Producer Election} Based on the statistics of the collected dataset, we find out that there are totally 2,009,168 accounts. Among these accounts, 1,739,839 accounts have staked EOS tokens as stakeholders while only 84,668 accounts have participated in the block producer election as voters, occupying 4.87\% of all stakeholders. The result indicates that only a small fraction of all stakeholders have taken part in the voting process. 

\begin{figure}[t]
	\centering
	\subfigure[Number evolution]{
		\label{distribution_number_voters_all_accounts}
		\includegraphics[scale=0.28, trim=0 0 0 0]{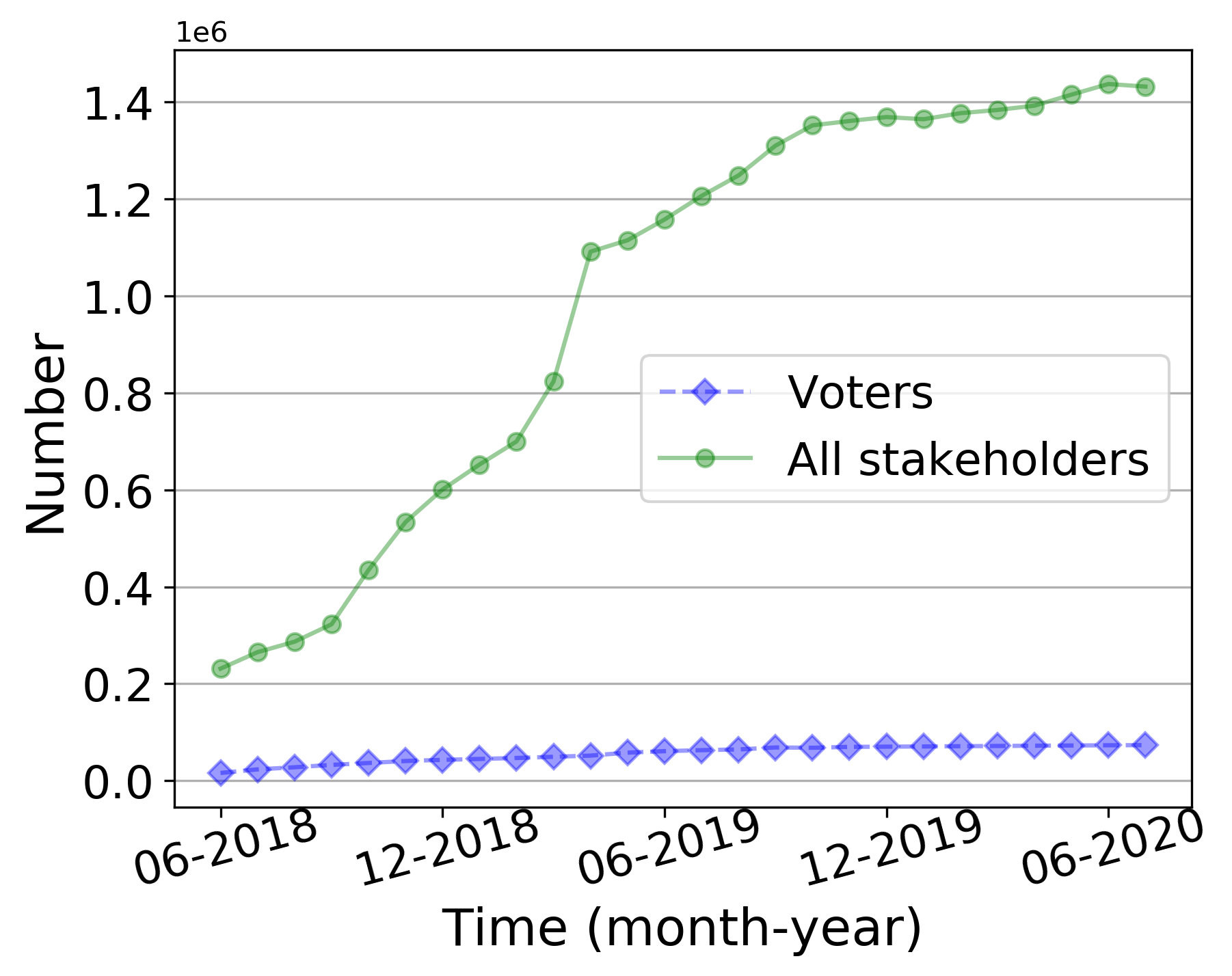}}
	\subfigure[Staked token evolution]{
		\label{distribution_voters_all_accounts}
		\includegraphics[scale=0.28, trim=0 0 0 0]{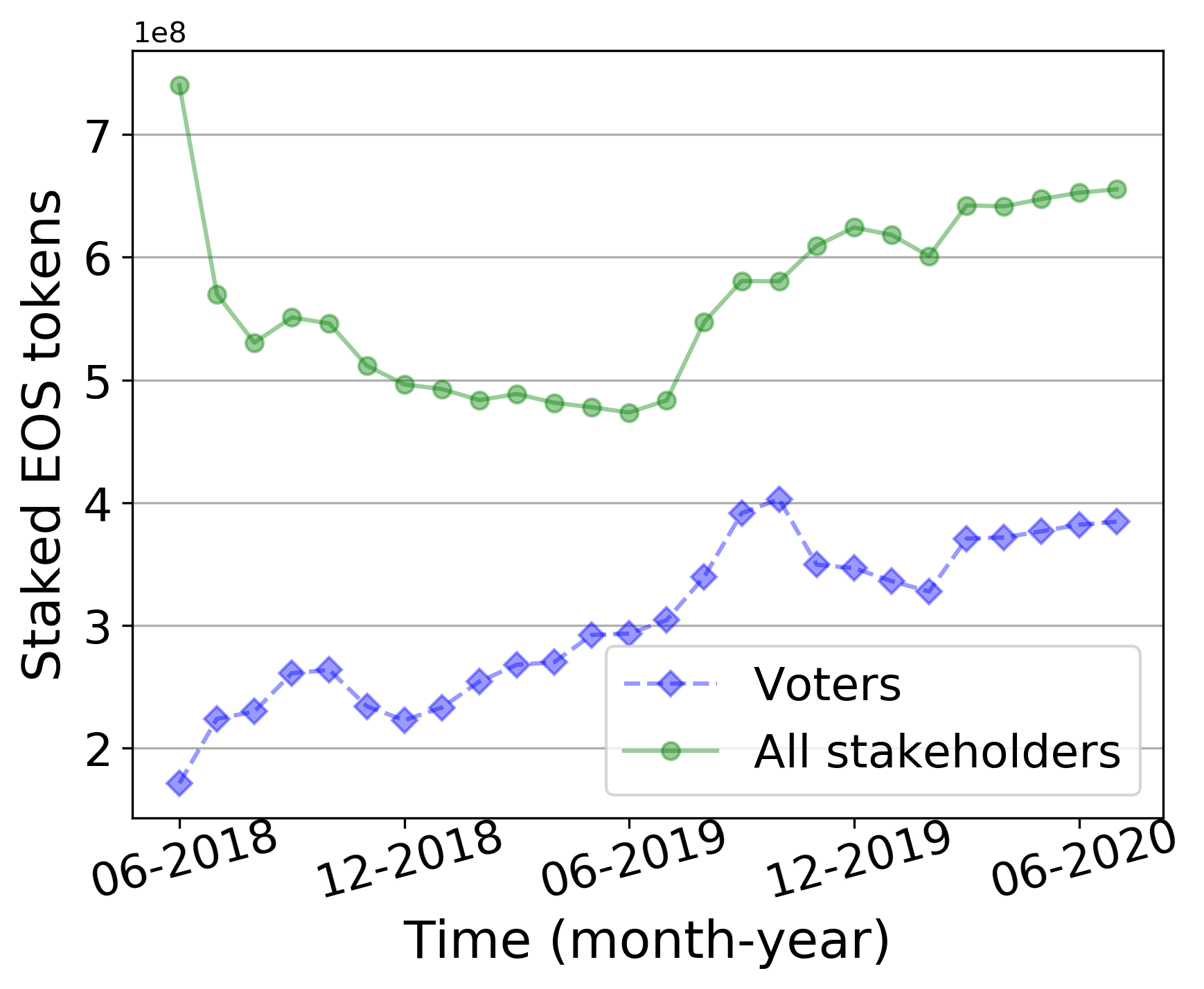}}
	\caption{The number evolution and staked token evolution of voters and all stakeholders.}
\end{figure}

We then calculate the number evolution of accounts for voters and all stakeholders, as shown in Fig. \ref{distribution_number_voters_all_accounts}. We found that the number of stakeholders and voters increase over time, but the number growth rate of voters is far short of the number growth rate of all stakeholders. This infers that many stakeholders do not care about the block production election, or even do not understand the voting mechanism in EOSIO. The power to decide who will become block producers thereby is held by a small number of stakeholders.

Besides, from the amount evolution of staked tokens held by voters and all accounts shown in Fig. \ref{distribution_voters_all_accounts}, we observe that the staked tokens held by voters account for half or less of the total network for a long time. Moreover, with the passage of time, the proportion of staked tokens held by voters increases. The waiver of trading fees in EOSIO provides a user-friendly environment, and thus DApp users will have less incentive to mortgage their EOS tokens. However, for application developers, voters, and block producer candidates, they have a strong motivation to mortgage their EOS tokens. Especially, block producer candidates always canvass other stakeholders for electoral support to ensure their position, and thus the amount of staked tokens of voters exhibits an increasing trend.

\subsubsection{Distribution Statistics}
\begin{figure}[t]
	\centering
	\subfigure[Staked token distribution of voters]{
		\label{voting_weight_distribution}
		\includegraphics[scale=0.28, trim=0 0 0 0]{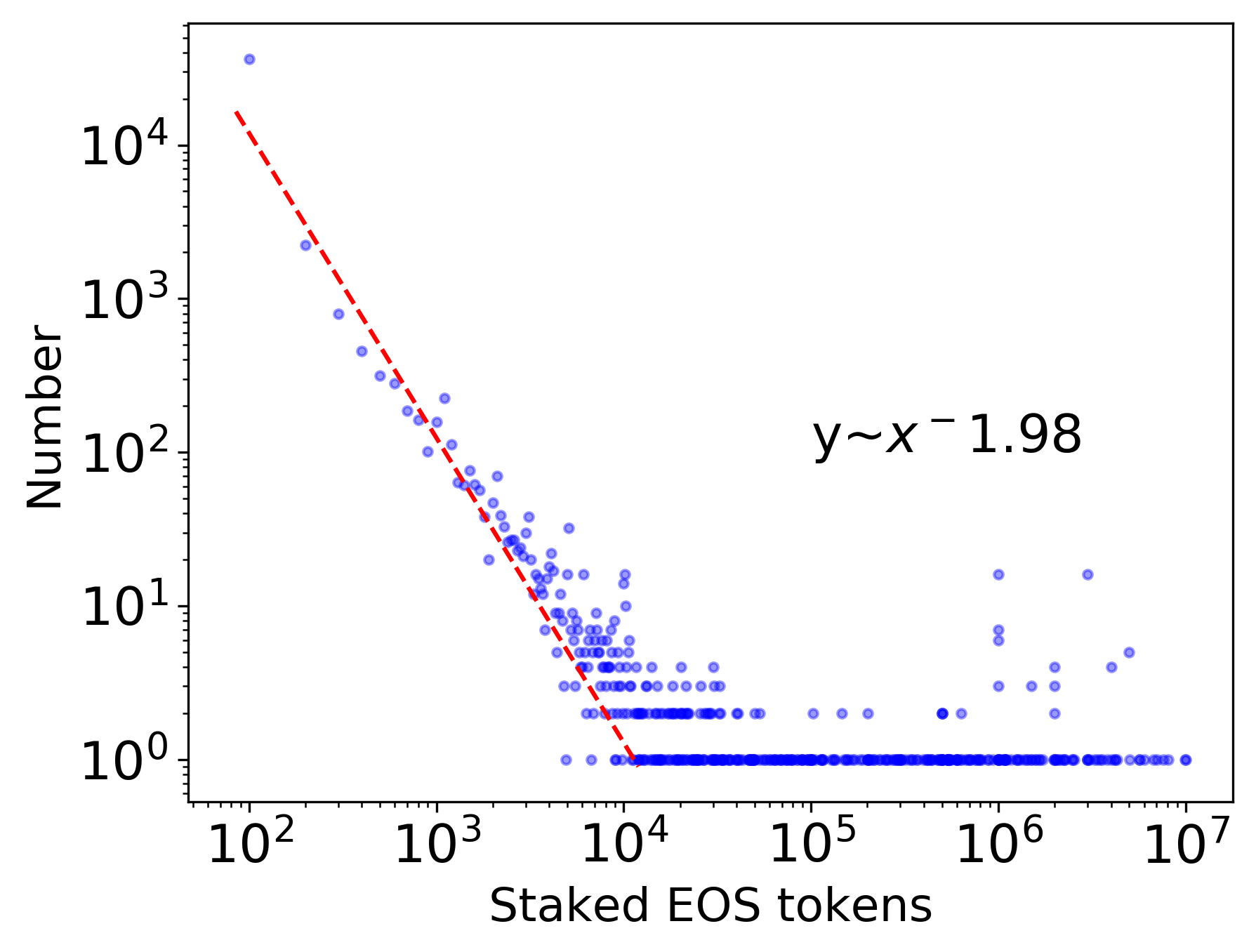}}
	\subfigure[Received voting weight distribution of block producer candidates]{
		\label{candidata_received_weight_distribution}
		\includegraphics[scale=0.28, trim=0 0 0 0]{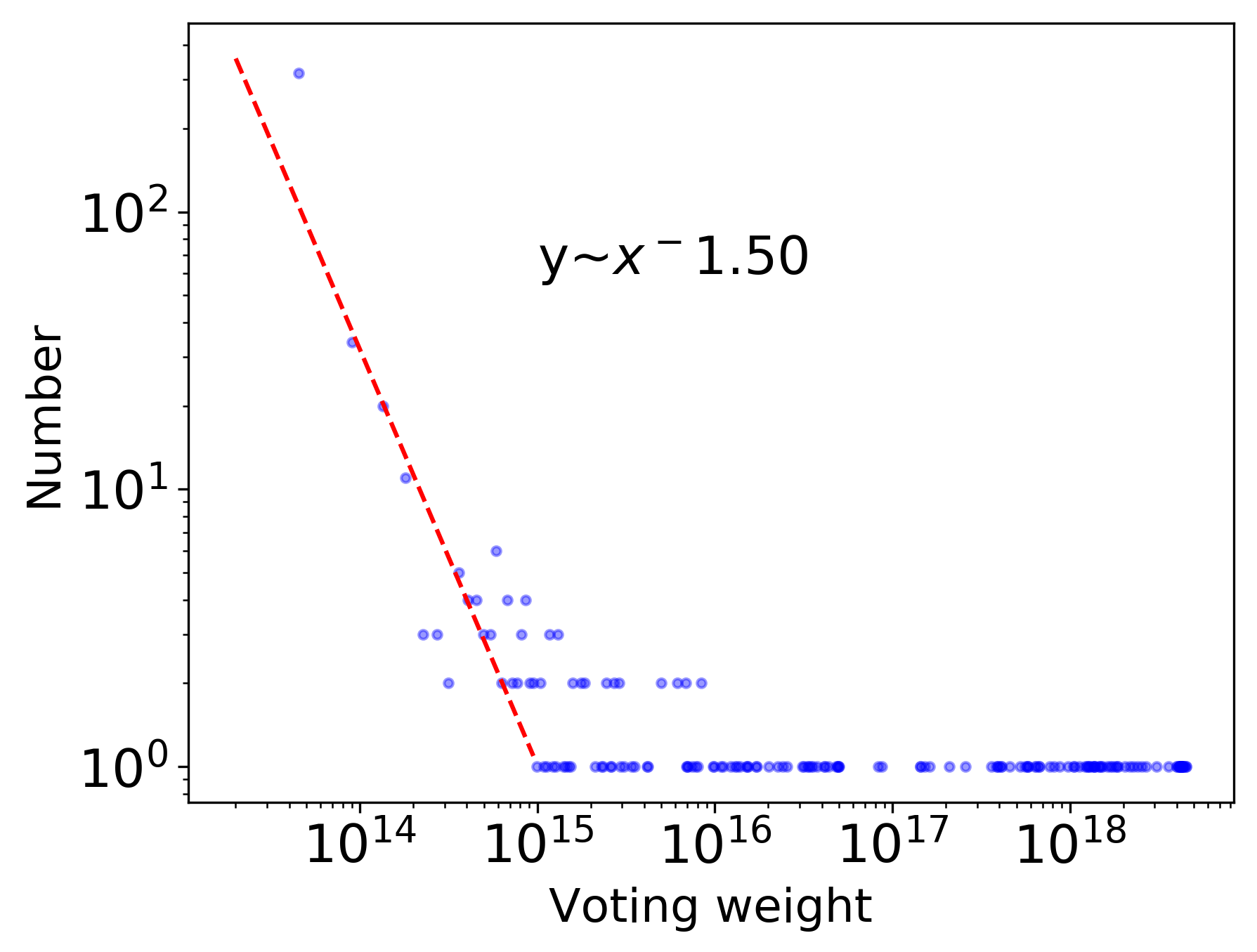}}
	\caption{The staked token distribution of voters (for proxies the tokens of proxied accounts are accumulated) and the received voting weight distribution of block producer candidates, calculated at the end of the 135,000,000 blocks.}
	\label{voting_distribution}
\end{figure}

We investigate the staked token distribution among all voters and the received voting weight distribution among all block producer candidates.
The distributions of the mapping values with a fitting line $y\sim x^{-\alpha}$ are shown in Fig. \ref{voting_weight_distribution} and Fig. \ref{candidata_received_weight_distribution}. Both of the two distributions are in line with the power-law distribution, which indicates that a small number of voters hold a large amount of staked tokens, and a few block producer candidates have received a large voting weight. We then calculate the fraction of stakes held by the richest voters, and find out that the top 5\% of the voters possess more than 95\% of the stakes. As for block producer candidates, the top 10\% of the candidates have received more than 95\% of the voting weight in the whole network. Hence, the voting resource is unevenly distributed in both voters and block producer candidates. As one of the security risks, most of the voting powers are concentrated on only a few rich voters so that the voting results basically depend on a small number of voters. On the other hand, the extremely uneven distribution of received voting weight among block producer candidates can lead to the rich-get-richer phenomenon, also named as the \textit{Matthew effect} \cite{perc2014matthew}. In the long run, it is easy to form block production monopoly among the most popular block producers.

\subsubsection{Statistics of Voting Proxies} Although anyone who stakes EOS tokens can participate in the governance of EOSIO with their voting power, block producer candidates should be carefully chosen by considering some factors such as resources, community contribution, and location of block producer candidates. However, not every stakeholder has enough incentive to investigate different candidates and stays abreast of every new proposal in EOSIO, especially those who only hold a tiny amount of EOS tokens.

Voting proxies are accounts that perform voting on behalf of other accounts. They vote for block producer candidates with voting power collected from proxied accounts, and communicate to the proxied accounts which block producer candidate they choose. Each user can delegate its voting power to a voting proxy, and can also withdraw its voting power from the proxy at any time. In general, voting proxies can remove the high barrier of participating in governance for users and fully arouse the initiative of individual users, thereby increasing user participation in the voting process.

\begin{figure*}[htbp]
	\centering
	\subfigure[Account number evolution]{\label{Statistic_proxy_voter_number}
		\includegraphics[scale=0.3, trim=0 0 0 0]{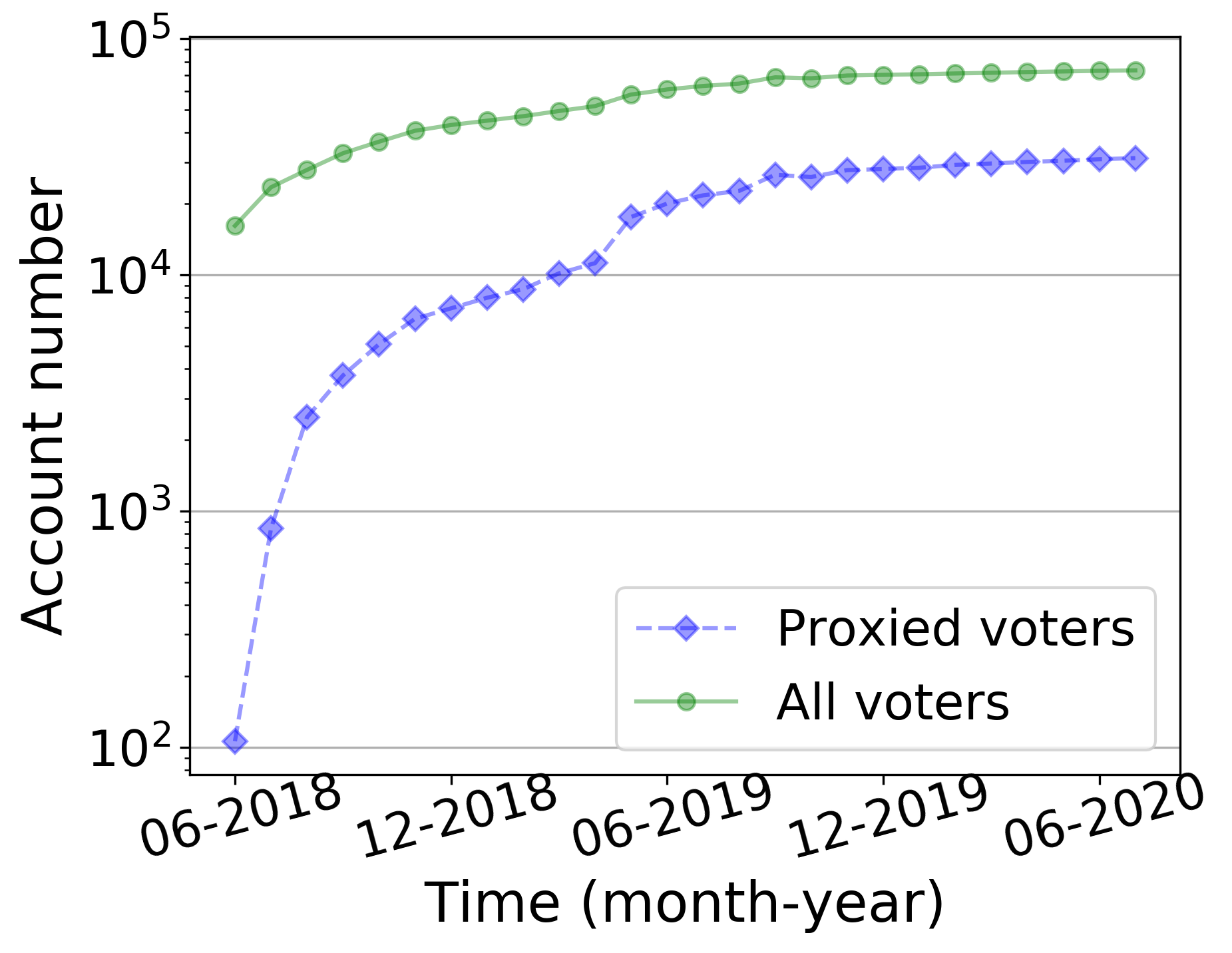}
	}
	\subfigure[Staked token evolution]{\label{Statistic_proxy_staked_token}
		\includegraphics[scale=0.3, trim=0 0 0 0]{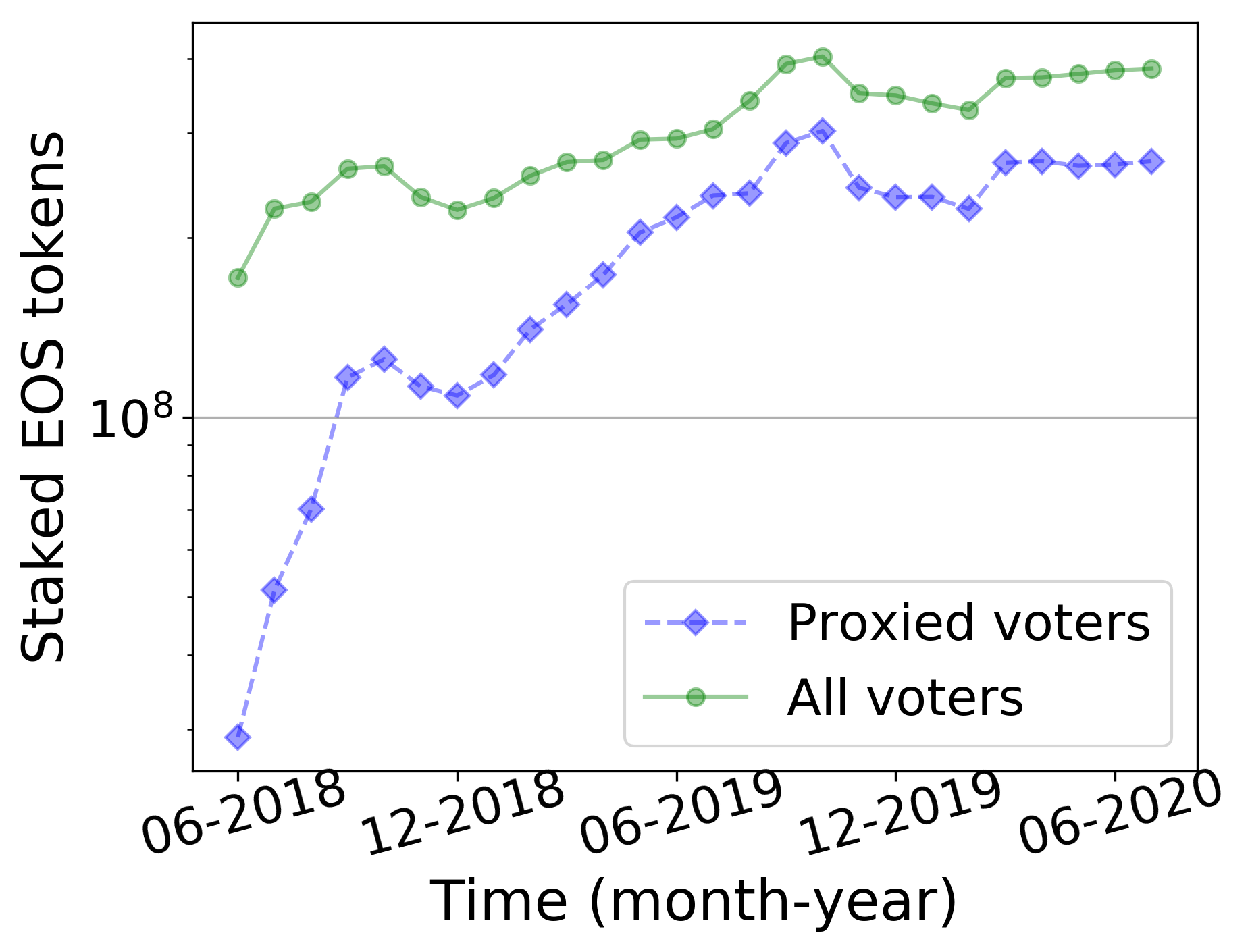}
	}	
	\subfigure[Voting weight evolution]{\label{Statistic_proxy_voting_weight}
		\includegraphics[scale=0.3, trim=0 0 0 0]{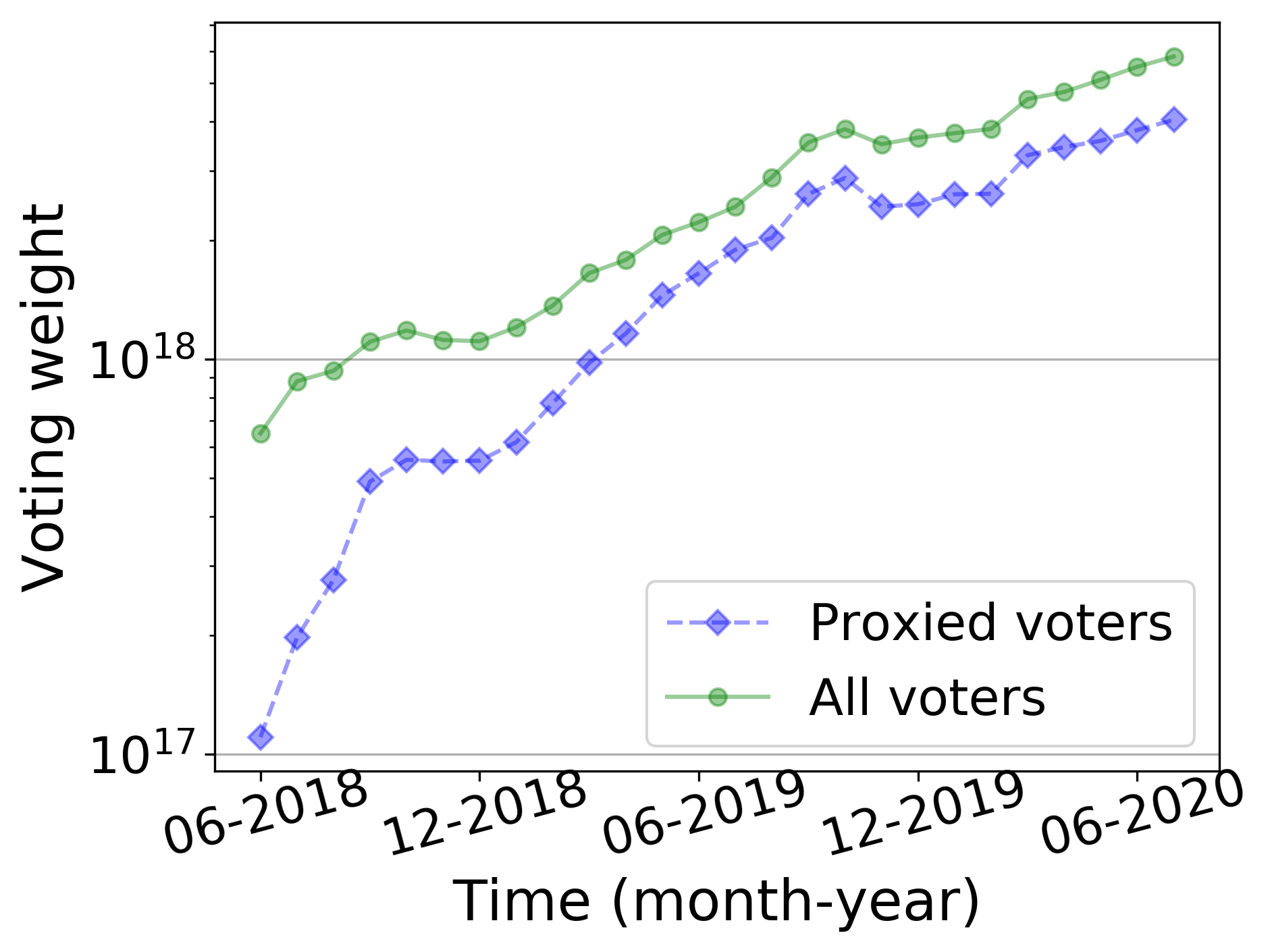}
	}
	
	\subfigure[Proportion evolution in account number]{\label{Statistic_proxy_voter_number2}
		\includegraphics[scale=0.3, trim=0 0 0 0]{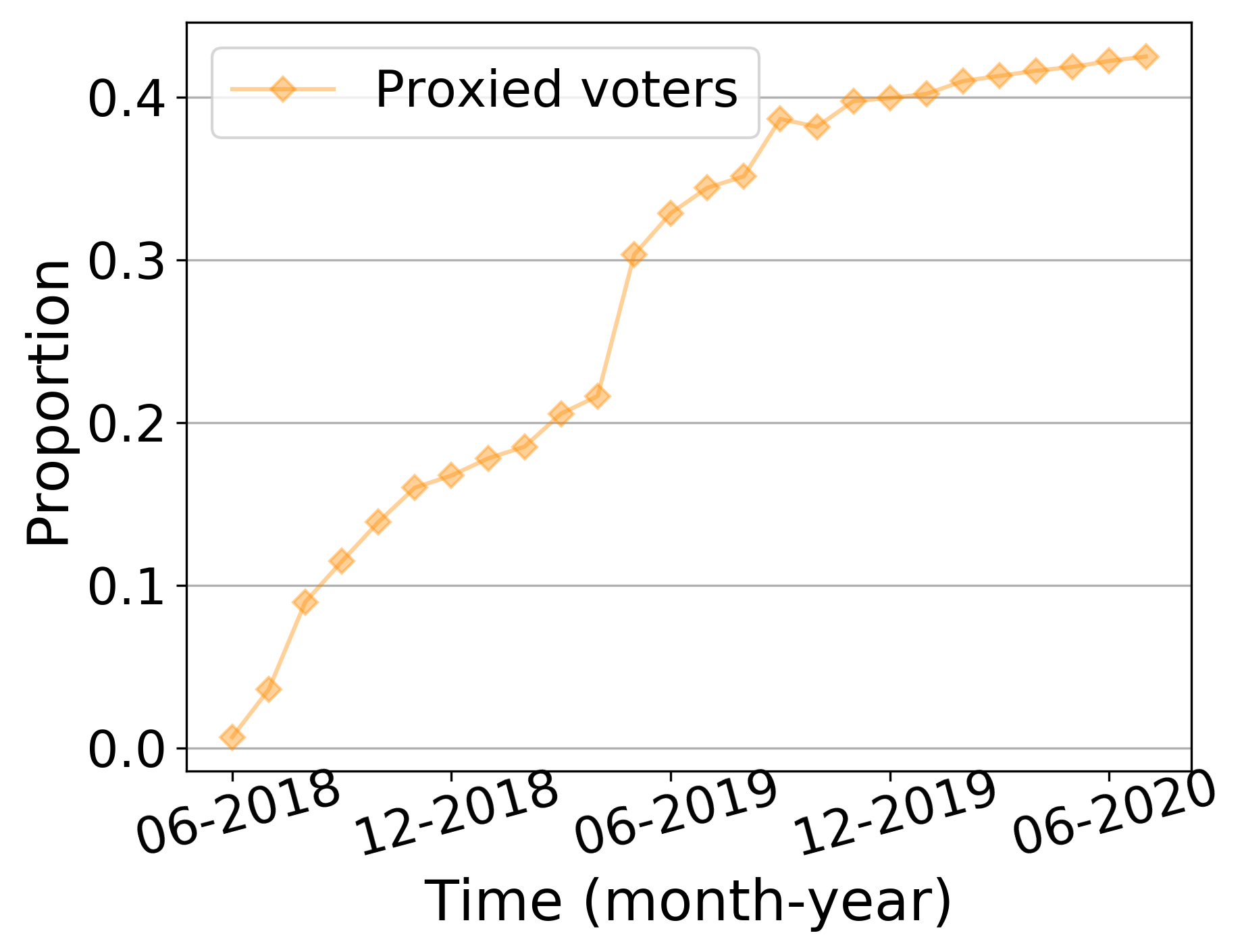}
	}
	\subfigure[Proportion evolution in staked tokens]{\label{Statistic_proxy_staked_token2}
		\includegraphics[scale=0.3, trim=0 0 0 0]{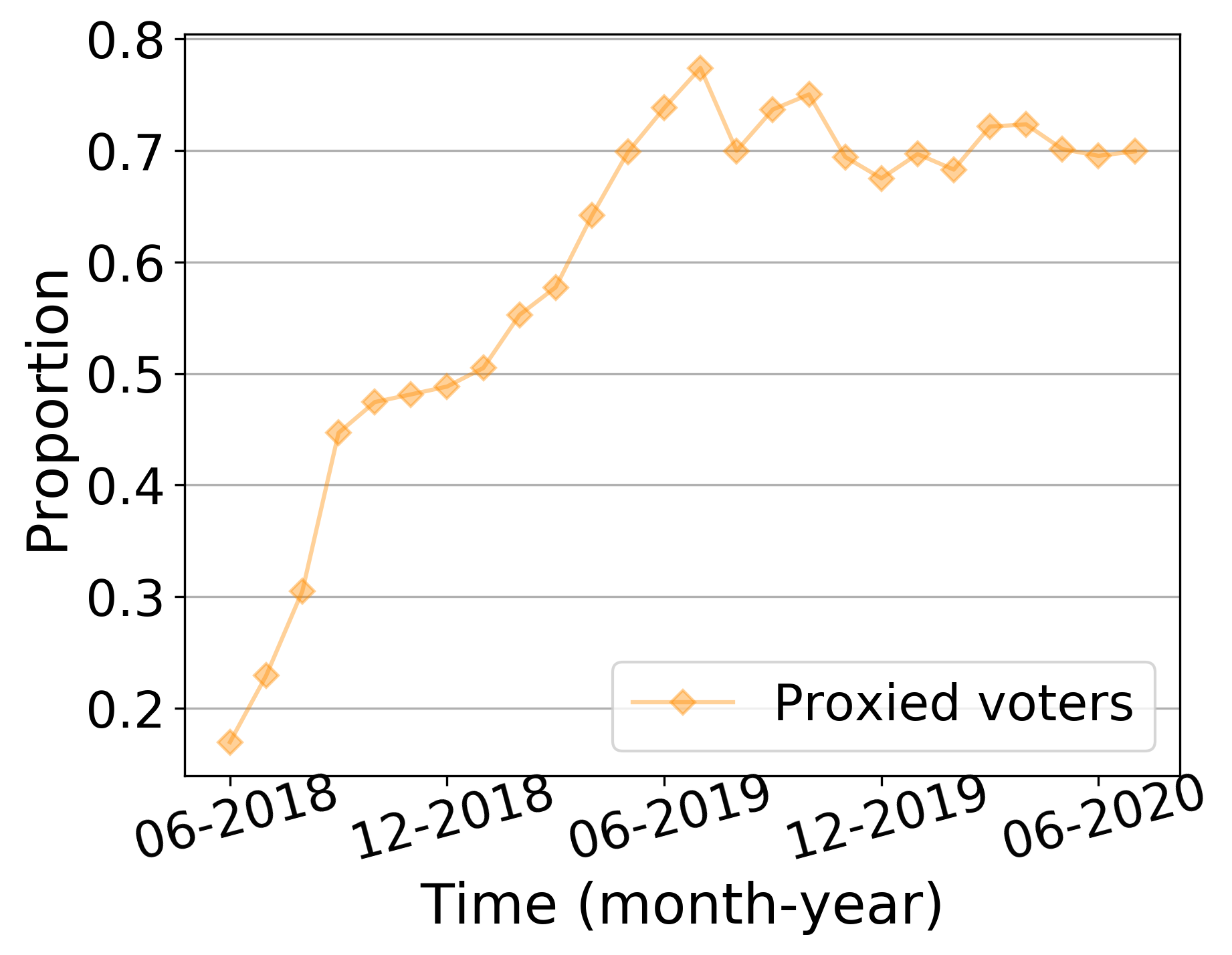}
	}	
	\subfigure[Proportion evolution in voting weights]{\label{Statistic_proxy_voting_weight2}
		\includegraphics[scale=0.3, trim=0 0 0 0]{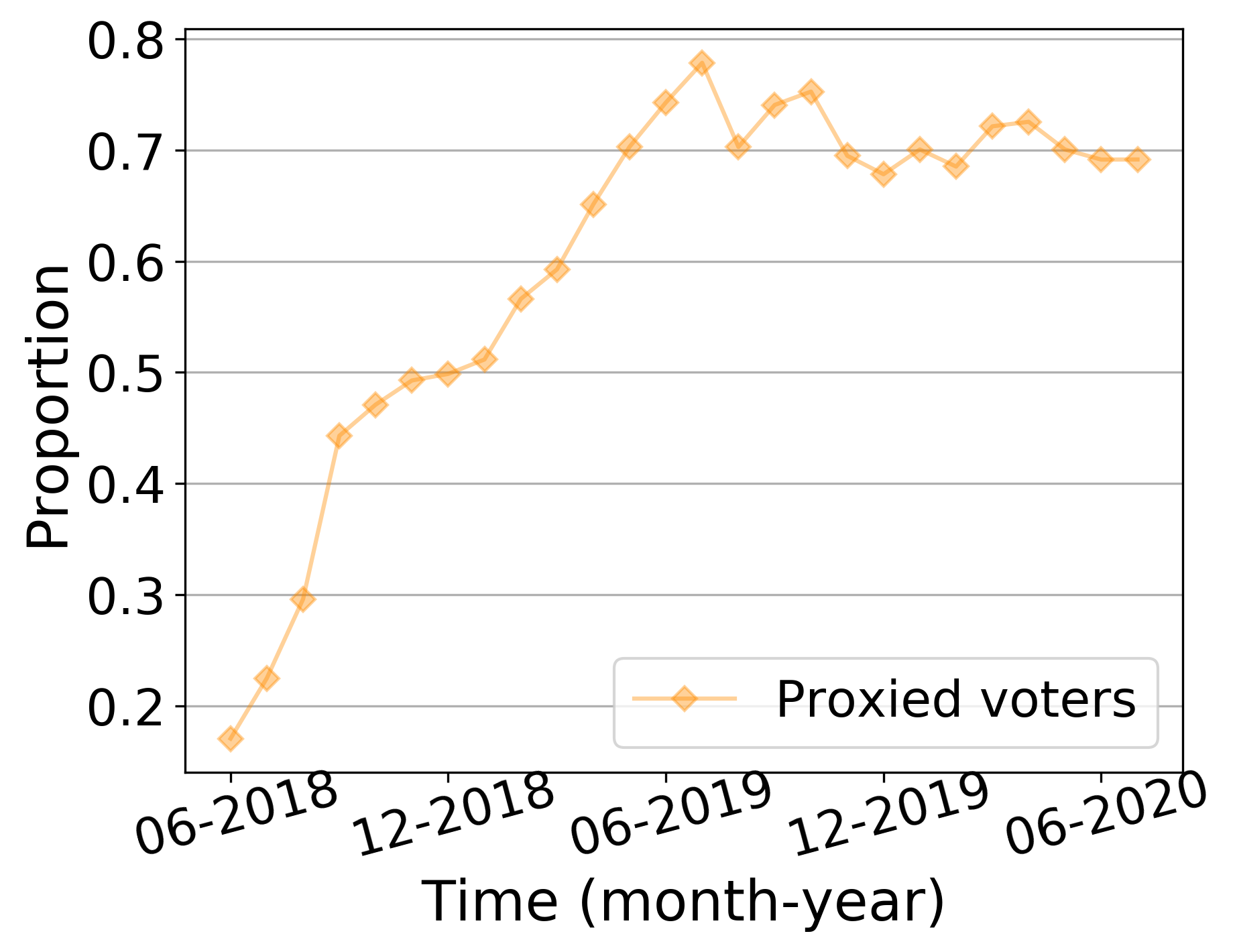}
	}
	\caption{Statistics about the proxied voters. (a)-(c) show the account number evolution, staked token evolution, and voting weight evolution of all voters and proxied voters. (d)-(f) display the value share evolution of the proxied voters in all voters for account number, staked tokens, and voting weights.}
	\label{fig_statistic_proxy}
\end{figure*}

According to our statistics, 1,792 accounts have registered to become voting proxies, and 924 of them have been delegated as proxies via the \textit{voteproducer} interface. It is worth noting that 40,561 of the 84,668 voters have voted through voting proxies, which means more than 47\% of the voters execute their voting power via voting proxies. Fig. \ref{Statistic_proxy_voter_number}-\ref{Statistic_proxy_voting_weight} show the account number evolution, staked token evolution, and voting weight evolution of all voters and proxied voters. By cooperating with the corresponding value share evolution of the proxied voters in all voters in Fig. \ref{Statistic_proxy_voter_number2}-\ref{Statistic_proxy_voting_weight2}, the increasing use of proxies can be observed. Moreover, after May 2019, nearly 70\% of the voting weights are delegated to voting proxies. Under this trend, the votes of voting proxies can exercise considerable influence over the election results, since most of the voting powers are centralized in the voting proxies. A valuable research issue is whether there exist abuses of rights among the voting proxies, and we will discuss this issue in Section \ref{VMD}.

\begin{figure}[t]
	\subfigure[Number evolution of block producers]{\label{fig_block_producer_number}
		\includegraphics[scale=0.27, trim=0 0 0 0]{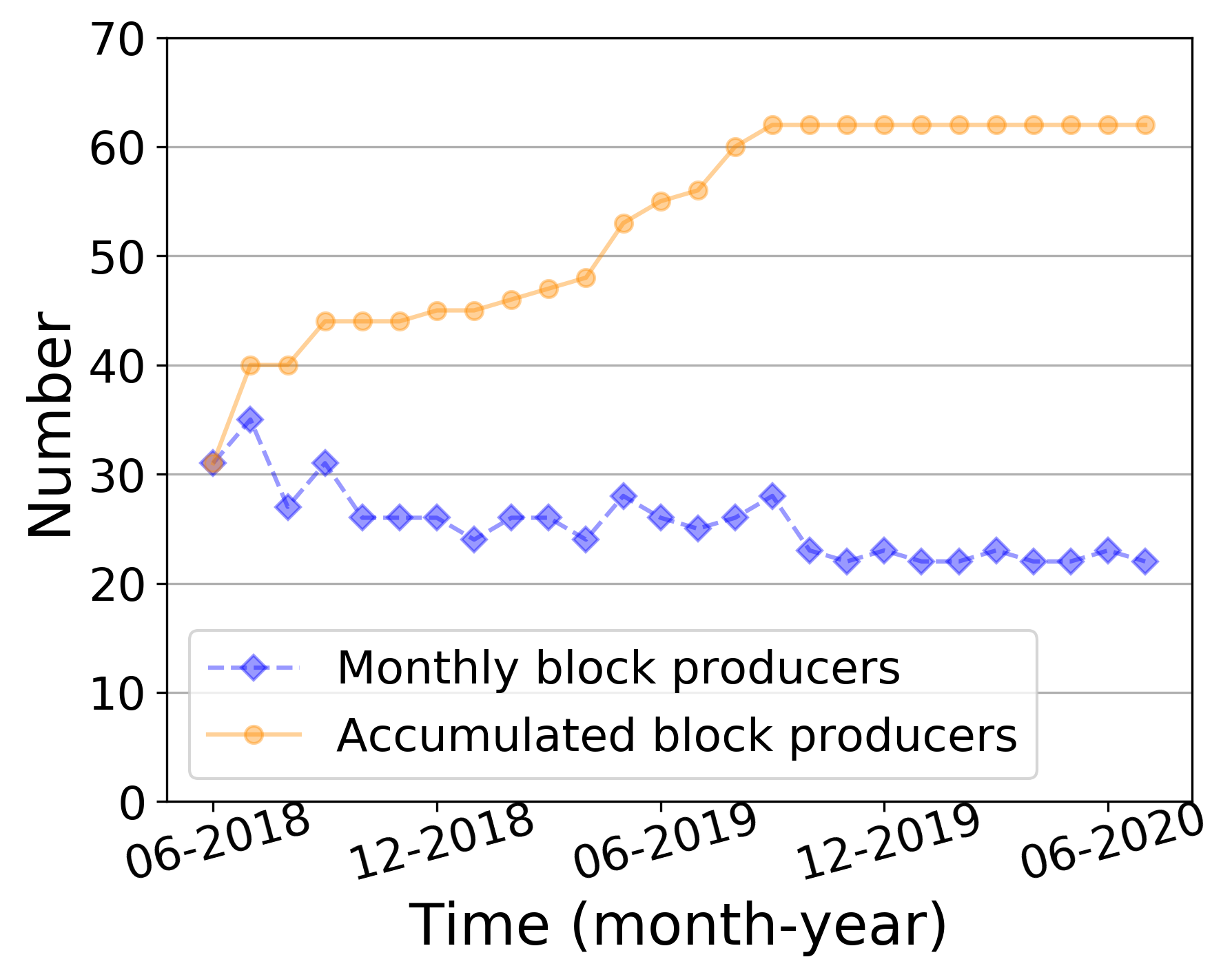}
	}
	\subfigure[Distribution of block producers' accumulative days participating in block production, calculated at the end of the 135,000,000 blocks]{\label{fig_block_producer_day}
		\includegraphics[scale=0.27, trim=0 0 0 0]{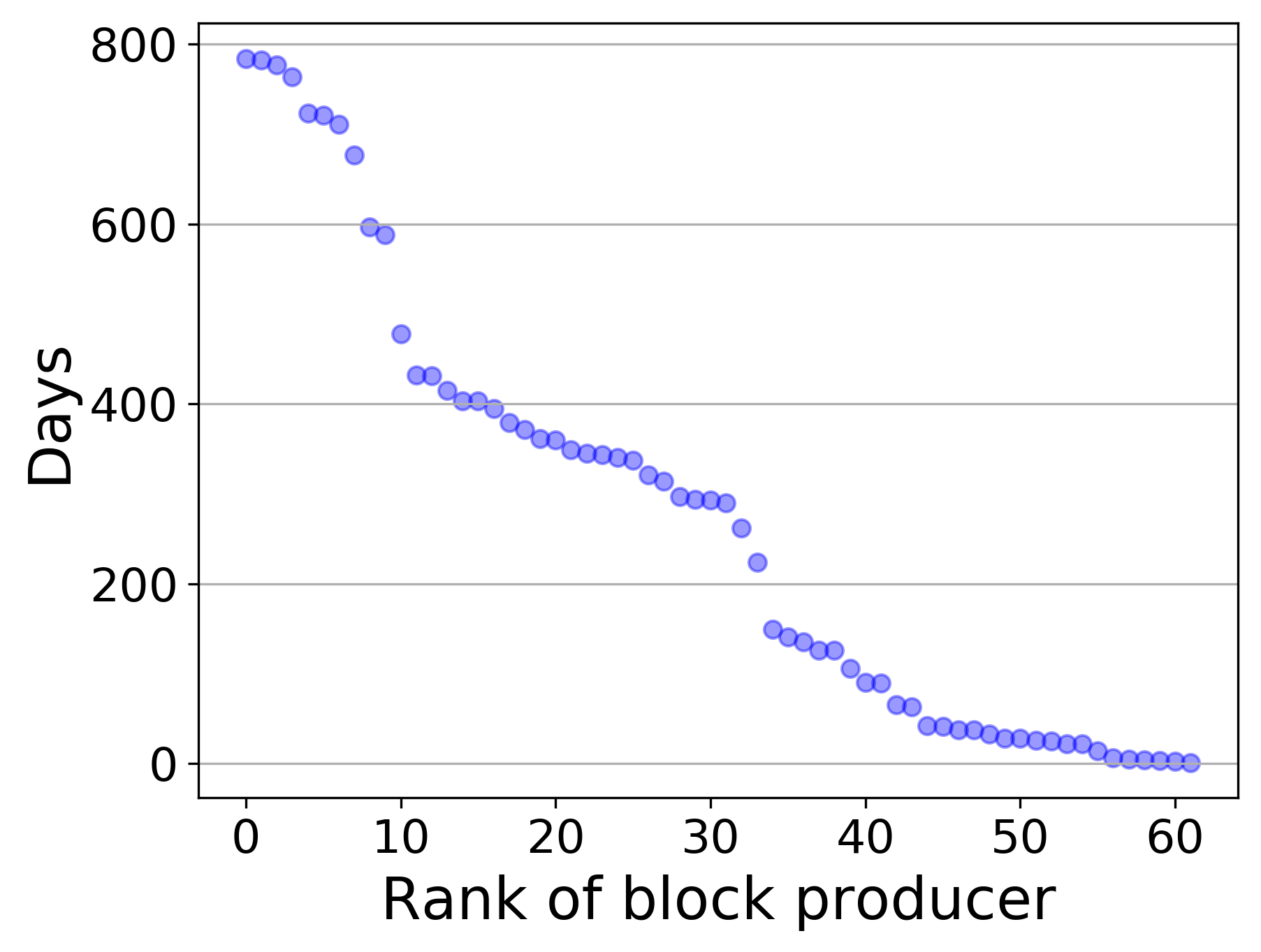}
	}	
	\centering
	\caption{Statistics about block producers, including (a) the number evolution and (b) distribution of days participating in block production.}
	\label{fig_block_producer_statistic}
\end{figure}
\begin{framed}
	\noindent \textbf{Finding 1:} The number of stakeholders increases rapidly with the boom of the EOSIO DApp ecology, but only a small number of them participated in the voting. The voting power is concentrated among a small number of voters.
	
	\noindent \textbf{Finding 2:} The received voting weights are unevenly distributed among the block producers, making it possible to form a block production monopoly in the future.
	
	\noindent \textbf{Finding 3:} Proxies account for an increasingly large proportion of the total voting power. Though the use of proxies can arouse more stakeholders to participate in the consensus government, it also centralizes most of the voting power in a small part of the accounts.
\end{framed}

\subsection{Block Production in EOSIO} \label{BPE}

\subsubsection{Overview of Block Production}
When focusing on block production, we found that there are 615 accounts registering to become block producer candidates.
Among these candidates, 602 candidates have received voting weights from the voters and 62 candidates have become block producers. Fig. \ref{fig_block_producer_number} shows the number evolution of block producers, and both the monthly number and accumulated number are given. Although the 21 elected block producers can be updated in each round according to their real-time voting weight, only around 25 different accounts are elected as block producers each month. Moreover, the chosen block producers are relatively fixed after September 2019. 

Fig. \ref{fig_block_producer_day} shows the distribution of days that block producers participating in block production, from which we can observe that about 10 block producers participating in block production for more than 600 days during the 135,000,000 blocks which cover about 790 days. These statistics illustrate that despite the large number of candidates, the set of block producers has small variations. Many accounts participate in block production for a long term. Once these accounts are in collusion, it is easy to achieve double-spend attacks since EOSIO can only tolerate no more than 1/3 of malicious block producers. In theory, the block producers acting in malicious manners can be voted out. Yet in practice, block producers can earn considerable rewards in block production so that most of them possess a large number of EOS tokens and can become one of the largest stakeholders, making it even harder to vote the malicious block producers out.   

\begin{figure*}[htbp]
	\centering
	\subfigure[Entropy with $n=10$]{\label{fig_entropy_top10}
		\includegraphics[scale=0.3]{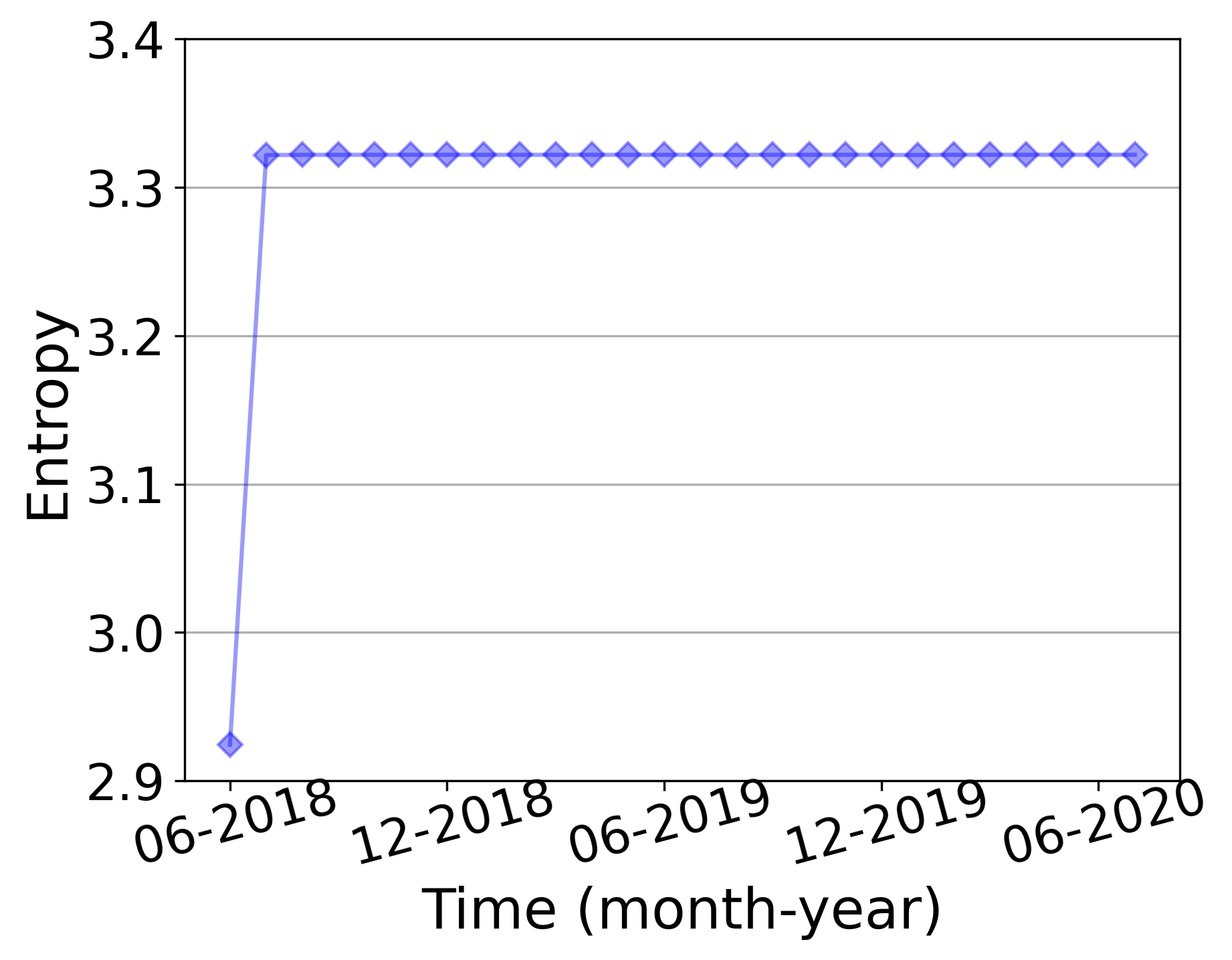}
	}
	\subfigure[Entropy with $n=20$]{\label{fig_entropy_top20}
		\includegraphics[scale=0.3]{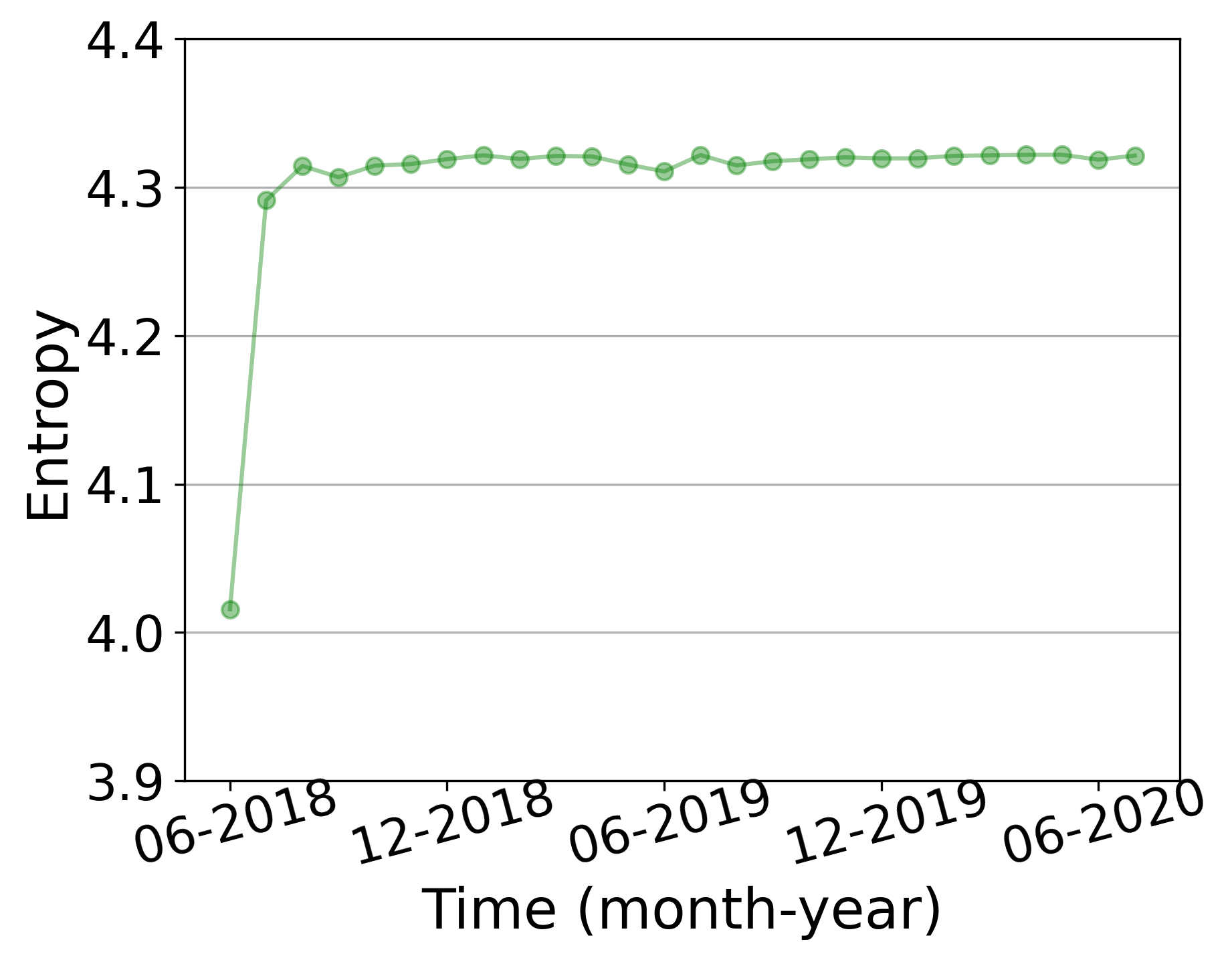}
	}
	\subfigure[Entropy with $n=$ the number of all block producers]{\label{fig_entropy_all}
		\includegraphics[scale=0.3]{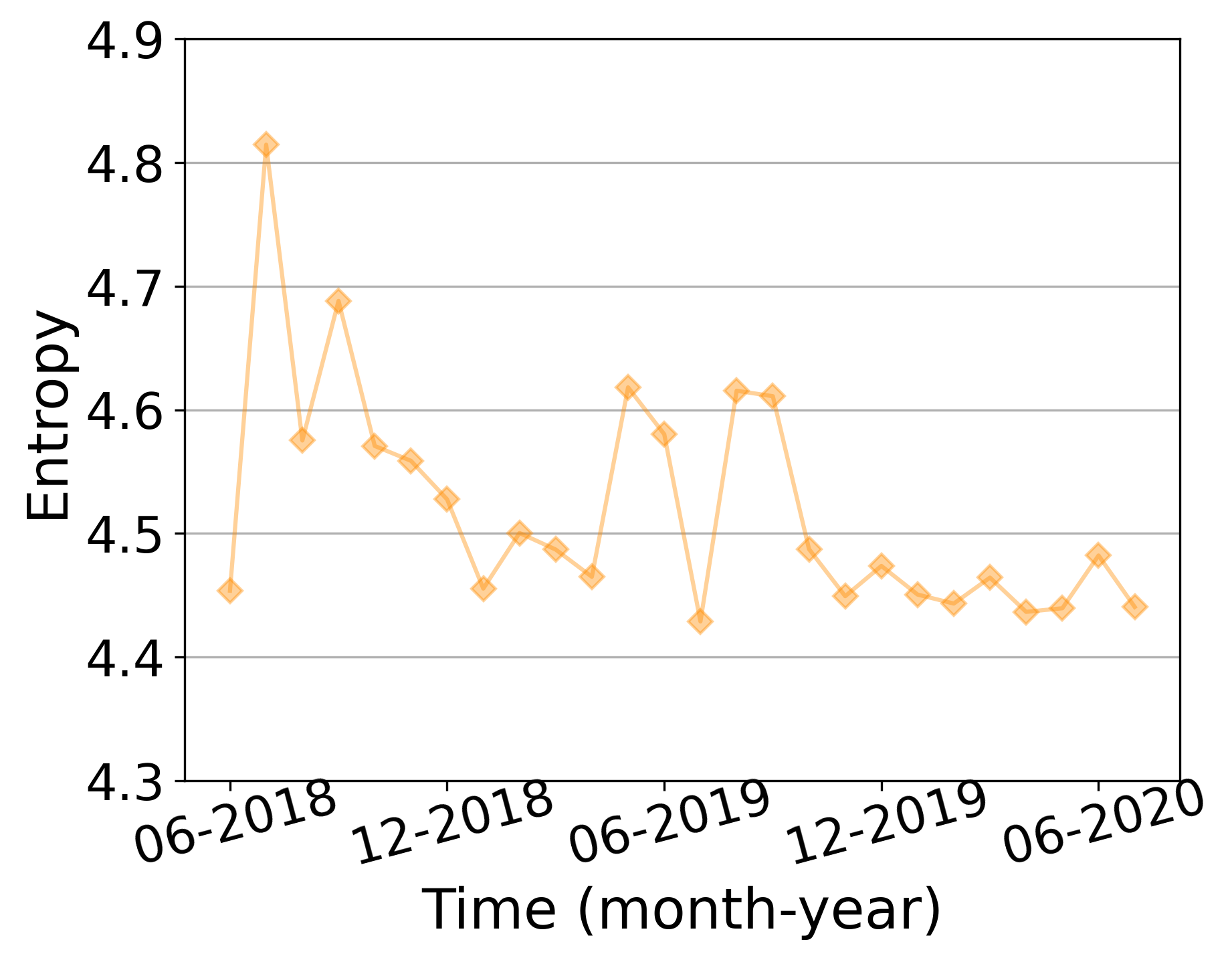}
	}
	\caption{Evolution of the information entropy.}
	\label{fig_entropy}
\end{figure*}

\subsubsection{Quantitative Analysis with Information Entropy} Previous studies have introduced using the information entropy metric to quantify the degree of decentralization in block production \cite{8784631,10.1007/978-3-030-59638-5_2}. Here we compare the decentralization degree of different months with information entropy. The information entropy metric for a month $i$, namely $H(X_i)$, can be calculated as follows:
\begin{equation}
p_j = \frac{x_j}{\sum_{j=1}^{n}x_j}, {\rm {where~}} x_j \in X_i,
\end{equation}
\begin{equation}
H(X_i) = -\sum_{j=1}^{n}p_jlog_2p_j,
\end{equation}
where $X_i$ stores the number of blocks produced by each producer in month $i$, $x_j$ denotes the number of blocks produced by block producer $j$, and $n$ denotes the measuring range of top-$n$ block producers. The greater the value of $H(X_i)$ is, the greater the decentralization degree is in the month, since the block production is more random and disorderly.

We display the quantitative results of information entropy in Fig. \ref{fig_entropy} by setting $n = 10, 20,$ and the number of block producers in each month. From Fig. \ref{fig_entropy_top10} and Fig. \ref{fig_entropy_top20}, we can observe that the information entropy of the first month is much smaller than that of the latter months for $n = 10$ and $n = 20$, which indicates that block production in the early days of EOSIO was more centralized among the top stakeholders. While as shown in Fig. \ref{fig_entropy_all}, block production in months Oct. 2018-Apr. 2019, Jul. 2019, Oct. 2019-Jul. 2020 were also relatively centralized in general. Combining with Fig. \ref{fig_block_producer_number}, we can infer that this phenomenon may be due to the less replacement of block producers in these months. 
\begin{framed}
	\noindent \textbf{Finding 4:} Despite the relatively large amount of candidates, the set of block producers has small variations in general.
	
	\noindent \textbf{Finding 5:} The block production was more centralized among the top stakeholders in the early days of EOSIO, and then it was more centralized in general for the later months due to the small variations of the block producer set.
\end{framed}

\section{Exploring the Voting Gang Anomalies}\label{VMD}
In the previous section, the voting election and block production in EOSIO are characterized statistically. The obtained observations have exposed some problems that may give rise to power centralization such as large stakeholders working together, proxies abusing their power, and block producers colluding. 
In this section, we identify and analyze the voting gang anomalies during the block producer election in EOSIO. Firstly, we develop a voter clustering method to detect those who potentially work together and share their voting targets. After that, we discuss the mutual voting behaviors in EOSIO and propose an anomaly detection method to identify suspicious voting gangs.

\subsection{Voter Clustering Analysis}
According to the statistics given in Section \ref{BPE}, there are totally 615 accounts registering as block producer candidates. Since each voter can vote for up to 30 block producer candidates and the votes can be updated in each round, the probability that two voters often vote for two groups of similar candidates is relatively low. Therefore, voters who often support similar block producer candidates are likely motivated by common interests. A great concern in EOSIO is that some large stakeholders may form alliances and communicate with each other before voting to solidify the positions of specific block producer candidates \cite{xu2018eos}, causing voting manipulation and unfair competition. Hence, we propose a clustering method based on the voting similarity of voters and analyze the clustering results.  
\subsubsection{Method Design}
\begin{algorithm}[t]\label{Algorithm}
	\caption{Voter Clustering Based on Similar Voting Behaviors}
	\begin{algorithmic}[1]
		\Require The set of voters $V$, voting records of voters $R$, selected threshold $\theta$.
		\Ensure A set of voter clusters with similar voting behaviors $C$.
		\State $visited = \emptyset$ $\verb|//|$visited marker
		\State $C = \emptyset$ $\verb|//|$cluster set
		\For{voter $v_i$ in $V-visited$}
		\State add $v_i$ to $visited$
		\State $C_i$ = \{$v_i$\}
		\State $N_{\theta} = \{v\in V,v!=v_i|$Similarity($R(v),R(v_i)$)$\geq\theta\}$
		\For{voter $v_j$ in $N_{\theta}$}
		\If{$v_j$ not in $visited$}
		\State add $v_j$ to $visited$
		\State add $v_j$ to $C_i$
		\For{voter $v_k\in \{v\in V,v!=v_j|Similarity(R(v),R(v_j)$)$\geq\theta\}$}
		\State add $v_k$ to $N_{\theta}$
		\EndFor
		\EndIf
		\EndFor
		\If{len($C_i$) != 1}
		\State add $C_i$ to $C$
		\EndIf
		\EndFor
	\end{algorithmic}
\end{algorithm}

To group the voters which have similar voting behaviors into a cluster, our method considers the similarity of the dynamic voting records of each voter. The approach contains two stages. Firstly, we collect the voting status of each voter every sample time, and represent the candidates voted by the voter into a set with a maximum length of 30. In the experiment, we record the voting status at the end of each month and thus we obtain 26 sets as the sampled voting records during Jun. 8, 2018 to Aug. 5, 2020 for each voter, which can be used in measuring the similarity of voters. Secondly, we propose a similarity-based voter clustering method which can automatically output the clusters with similar voting behaviors. Algorithm 1 shows the pseudocode of the method, where the input contains the set of voters $V$, the sampled voting records $R$, and a similarity threshold $\theta$. The set $visited$ in the algorithm is used to mark the visited voters, and we treat each unvisited voter as a center. For each center, we select out voters whose voting similarity is greater than or equal to $\theta$, add them to $visited$ and the cluster that the center belongs to. And then, if a center $v_i$ has members behaving similarly in its cluster $C_i$, voters who are similar to these members are also added into $C_i$. Finally, the cluster $C_i$ is added to the output cluster set $C$. 
Technically speaking, the proposed method detects voter clusters based on the similarity of the dynamic voting records, and the similarity measurement can be Jaccard's coefficient, Pearson correlation coefficient, etc.

\subsubsection{Detecting in Large Stakeholders}


\begin{figure*}[t]
	\centering
	\subfigure[Three voting snapshots]{\label{cluster_large_stakeholder}
		\includegraphics[scale=0.36]{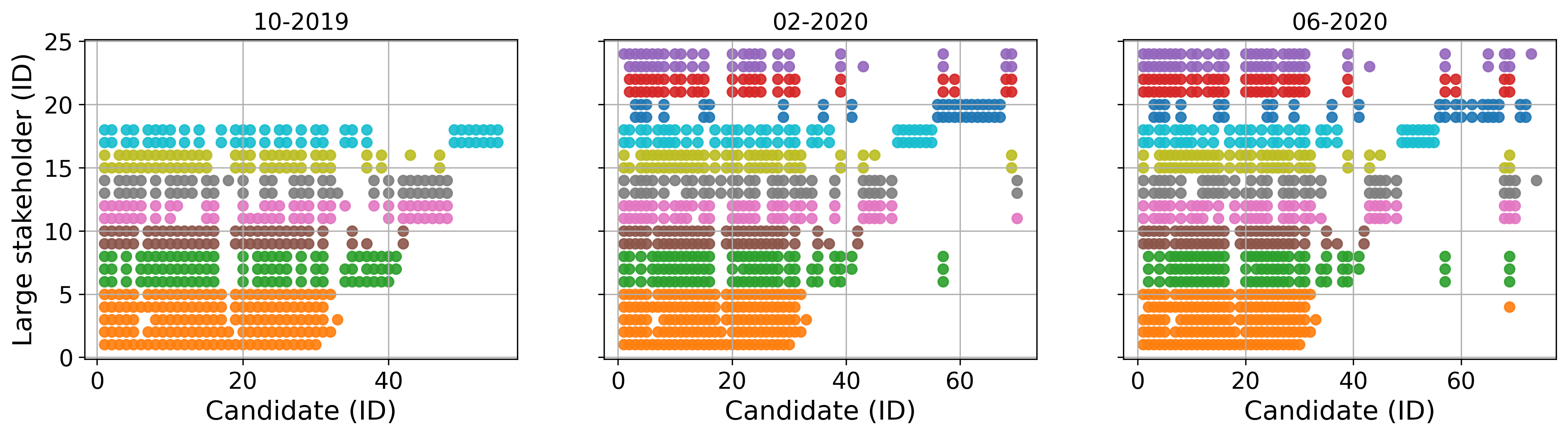}
	}
	\subfigure[Voting time distribution]{\label{cluster_time_largestakeholder}
		\includegraphics[scale=0.36]{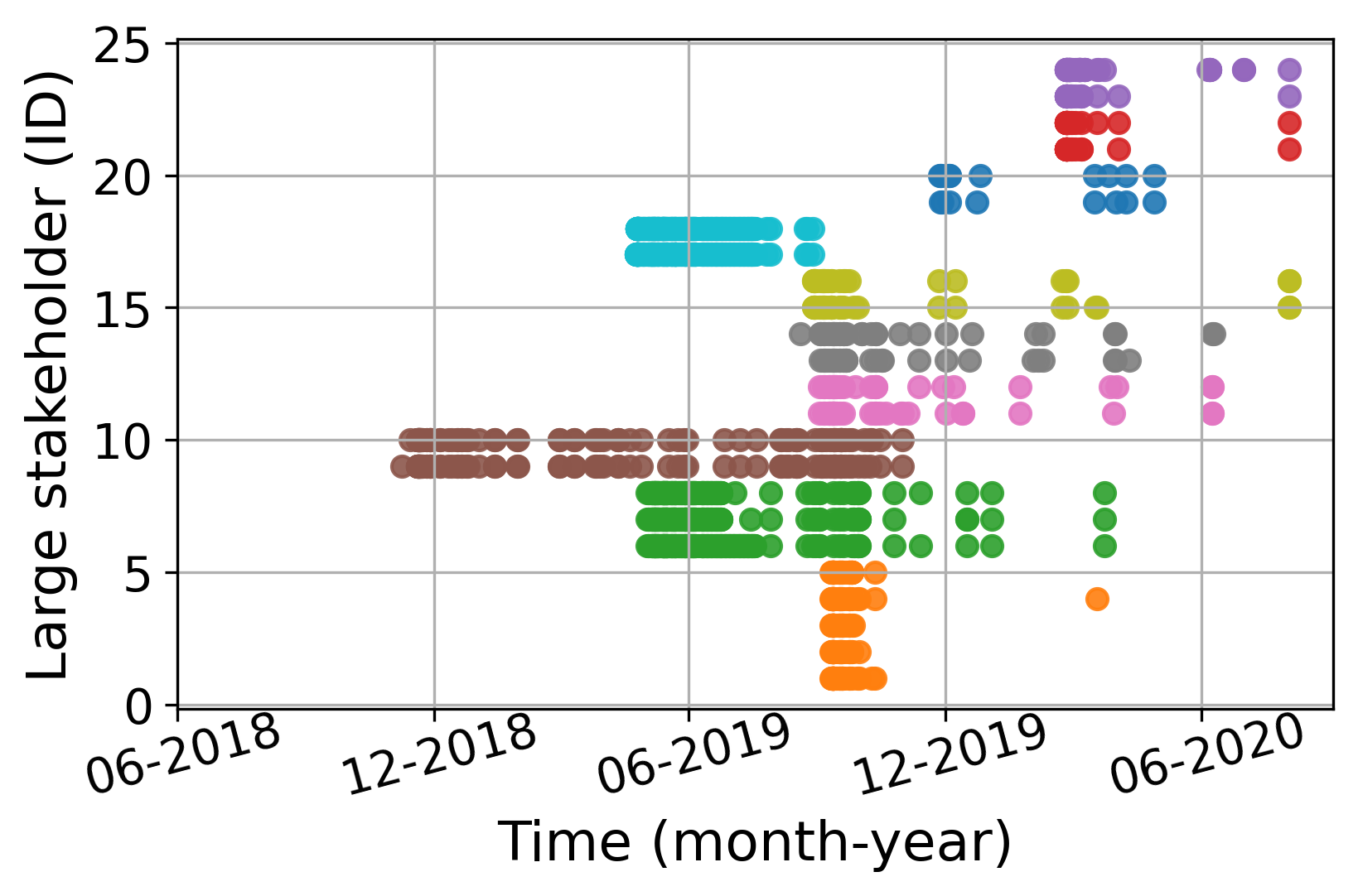}
	}
	\caption{Three voting snapshots and the voting time distribution of the top 10 detected clusters with the most staked tokens among the large stakeholders, where votes in the same cluster are assigned the same color.}
	\label{fig_large_stakeholder_clusters}
\end{figure*}



\begin{figure}[t]
    \centering
	\subfigure[Large stakeholders]{\label{cluster_creator_largestakeholder}
		\includegraphics[scale=0.14]{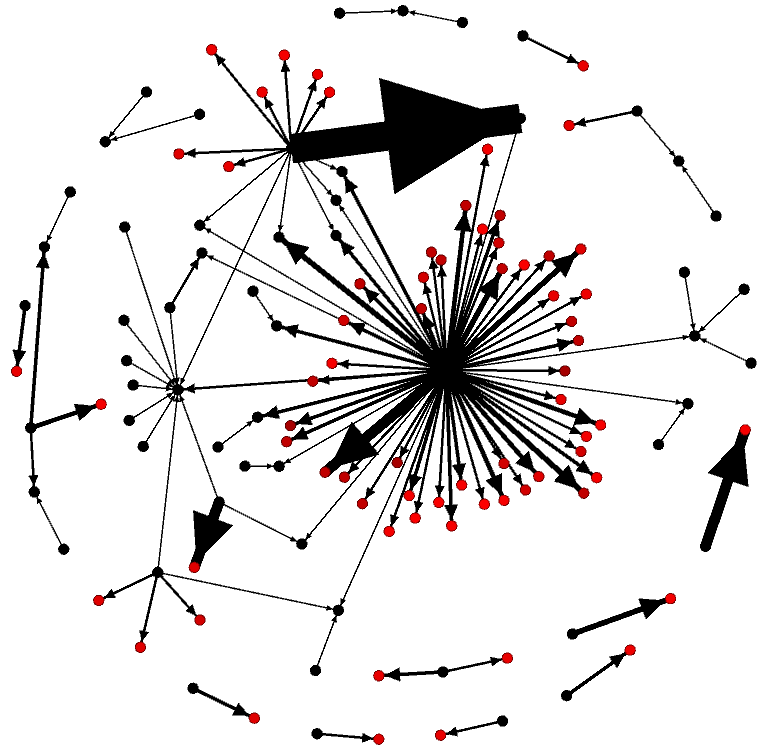}
	}
	\subfigure[Proxies]{\label{cluster_creator_proxies}
		\includegraphics[scale=0.156]{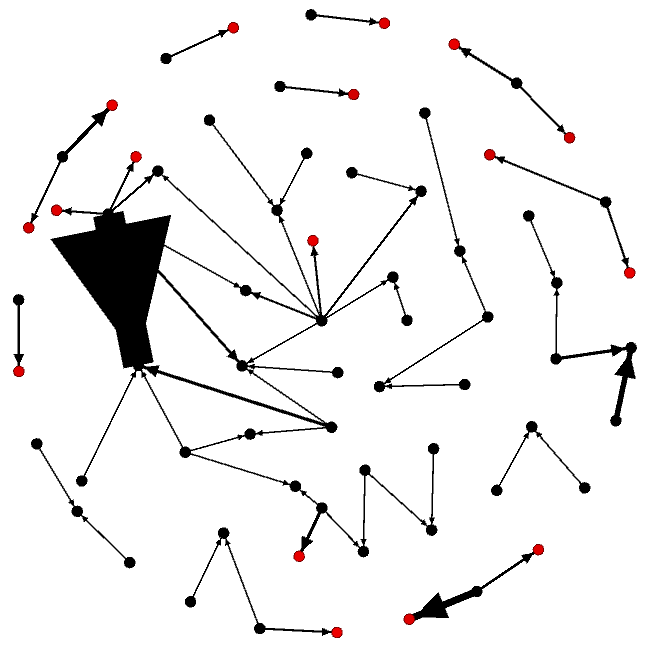}
	}
	\caption{Relationships between account creators and the detected clusters, where clusters owning one account creator are colored red.}
\end{figure}

As mentioned in Section \ref{BPEE}, the staked token distribution among all voters is extremely uneven. About 5\% of the voters possess more than 95\% of the stakes in the statistics. Hence the votes of large stakeholders make a great effect on the governance of EOSIO. 
To investigate the large stakeholders with similar voting behaviors, we apply Algorithm 1 on the top 5\% richest voters (the proxied stakes are accumulated for proxies when ranking) in the sampled voting records, consisting of 2,723 voters. With Jaccard's coefficient as the similarity measure and $\theta = 0.9$, we detect 88 clusters including 293 voters.

Fig. \ref{cluster_large_stakeholder} displays the votes of the large stakeholders in the top 10 detected clusters with the most staked tokens, where the y-axis depicts the index of the detected large stakeholders and the x-axis depicts the index of the candidates. If a voter $y$ votes for block producer candidates $x_1, x_2$ and $x_3$, then ($x_1$, $y$), ($x_2$, $y$) and ($x_3$, $y$) will be colored in the corresponding color of its cluster. 
From this figure, we can observe that voters in the same detected cluster have similar voting behaviors in both time and space, which can be illustrated by the very similar block producer candidates they support in these different snapshots.
Fig. \ref{cluster_time_largestakeholder} shows the distribution of time when the large stakeholders in the top 10 detected clusters with the most staked tokens operate the \textit{voteproducer} action, where different clusters are assigned different colors. To our surprise, though the voters in the same cluster are grouped according to the sampled voting records, the voting actions of these voters occur at very similar timestamp sequences.

We further analyze from the account creators of the detected voters, and visualize the relationships between account creators and the 88 clusters in Fig. \ref{cluster_creator_largestakeholder}. In each edge of the figure, a target node represents a detected cluster of large stakeholders, a source node represents an account creator that creates one or more stakeholders in a cluster, and the edge thickness is related to the number of accounts created. From the figure, we can observe that most accounts in the same cluster are created by the same account, implying that these voters are likely to be controlled by the same entity and therefore they have similar voting behaviors. Such kind of gathering phenomena in voting among large stakeholders can easily cause block production monopoly\textemdash the allies of these large stakeholders can obtain a solid position in the election. Moreover, they will earn more rewards (i.e, EOS tokens) from block production to consolidate their advantage in the election.

\subsubsection{Detecting in Voting Proxies}


\begin{figure*}[t]
	\centering
	\subfigure[Three voting snapshots]{\label{cluster_proxy}
		\includegraphics[scale=0.36]{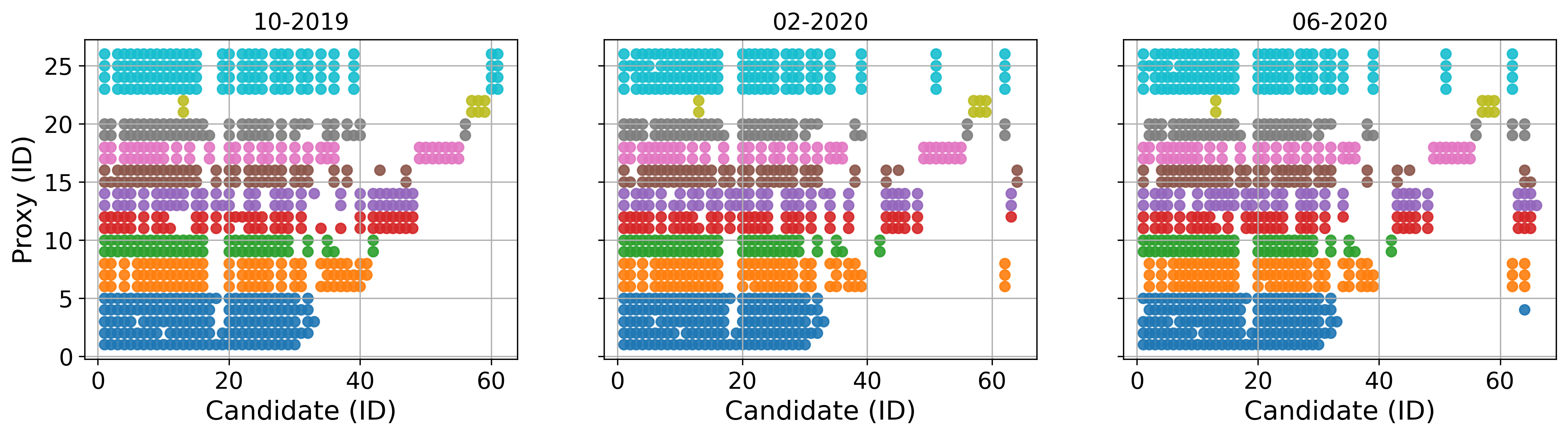}
	}
	\subfigure[Voting time distribution]{\label{cluster_time_proxy}
		\includegraphics[scale=0.36]{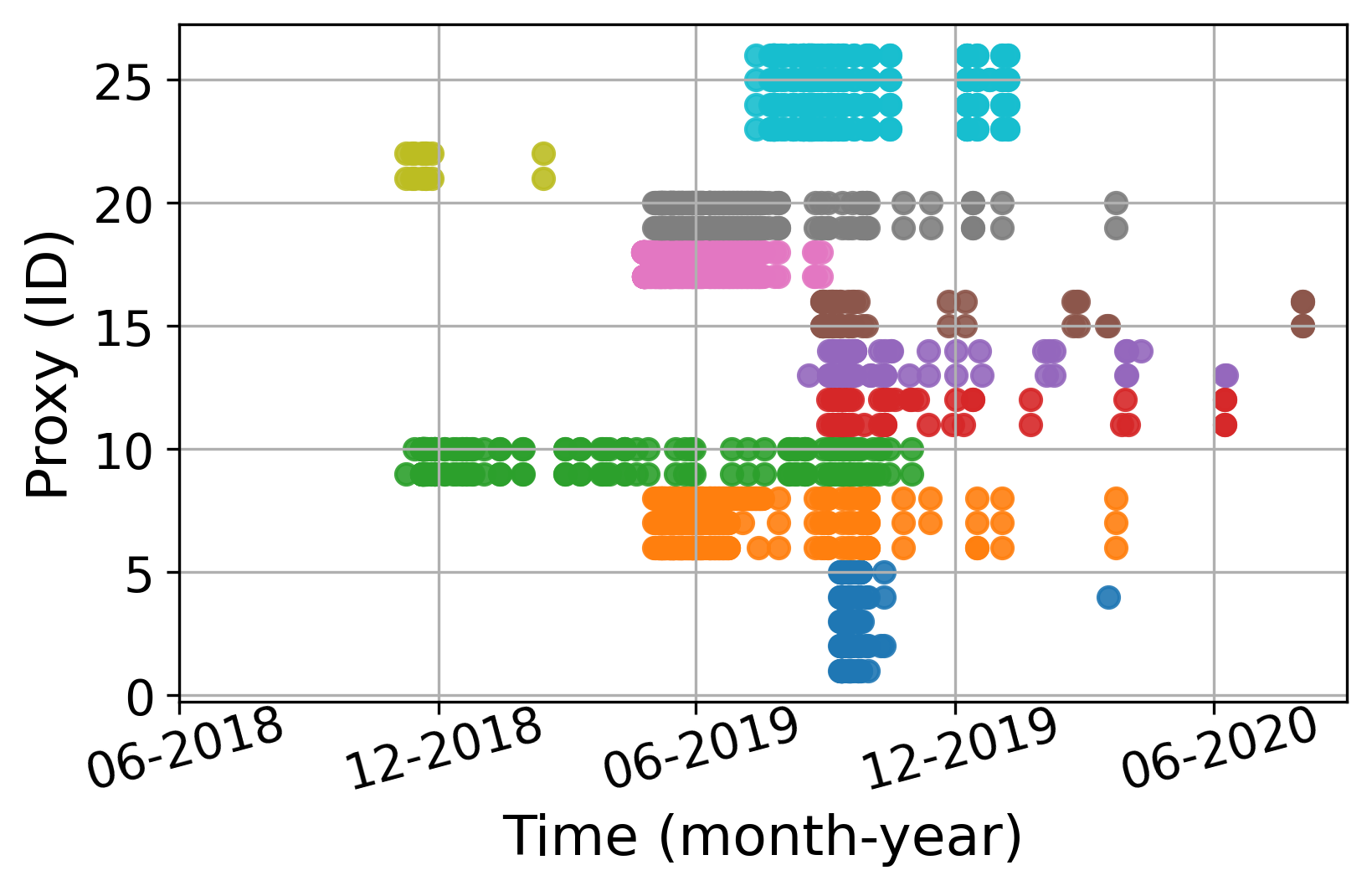}
	}
	\caption{Three voting snapshots and the voting time distribution of the top 10 detected clusters with the most staked tokens among the proxies, where votes in the same cluster are assigned the same color.}
	\label{fig_proxy_clusters}
\end{figure*}

Voting proxies are accounts that can execute the voting power of proxied accounts. Recently, we have seen the rising use of proxies in voting. Usually, users delegate their voting rights to a proxy out of trust in the proxy's voting decision. However, we can not exclude the presence of ``selfish'' proxies that only vote for their alliance members but do not consider the governance of the community. In this way, if users delegate their voting power to these proxies, they will actually increase the competitiveness of a particular alliance and the degree of centralization. Therefore, there is an urgent need to understand the voting behaviors of voting proxies. Particularly, proxies having similar voting behaviors deserve our attention. And these proxies are suspected of being manipulated by the same entity and using different identities to attract the authorization of other stakeholders.

To this end, we apply Algorithm 1 on all proxies in our dataset with Jaccard's coefficient as the similarity measure and $\theta=0.9$. We totally detect 35 clusters including 162 proxies. Fig. \ref{cluster_proxy} and Fig. \ref{cluster_time_proxy} visualize three voting snapshots and the voting time distribution of proxies in the top 10 detected clusters with the most staked tokens, where different clusters are assigned different colors. Similar to the observations drawn from Fig. \ref{cluster_large_stakeholder} and Fig. \ref{cluster_time_largestakeholder}, the proxies in the same detected clusters vote for almost the same group of block producer candidates across time, and the time they operate the \textit{voteproducer} action shows a convergent trait.

Fig. \ref{cluster_creator_proxies} displays the account creation relationships between account creators and the 35 detected clusters, and a target node denotes a cluster of voting proxies and a source node denotes an account creator that creates one or more proxies in a cluster. As we can see, among 17 of the 35 clusters, proxies in the same cluster are created by one account. Besides, we observe that many proxies in the same cluster have common features in their account name. For example, proxies \textit{`hashfineos44'}, \textit{`hashfineos14'}, \textit{`hashfineos33'}, \textit{`hashfineos55'} and \textit{`hashfineos13'} have the common prefix `hashfineos-' in their name, proxies \textit{`reallyreally'}, \textit{`windowwindow'}, \textit{`citycitycity'}, \textit{`familyfamily'}, etc. have reduplicated words in their name. We also observe that the account \textit{`octgenerator'} creates 53 proxies detected in the same cluster, and all these proxies only vote for \textit{`oraclegogogo'}. Though it is impossible to verify that the detected accounts in the same cluster belong to an entity due to the pseudonymous nature, our method helps reveal the abnormal voters who potentially work together.


\subsection{Mutual Voting Anomaly Detection} 


\begin{figure}[t]
	\centering
	\subfigure[Mutual voting patterns]{\label{mutual_voting_pattern}
		\includegraphics[scale=0.26]{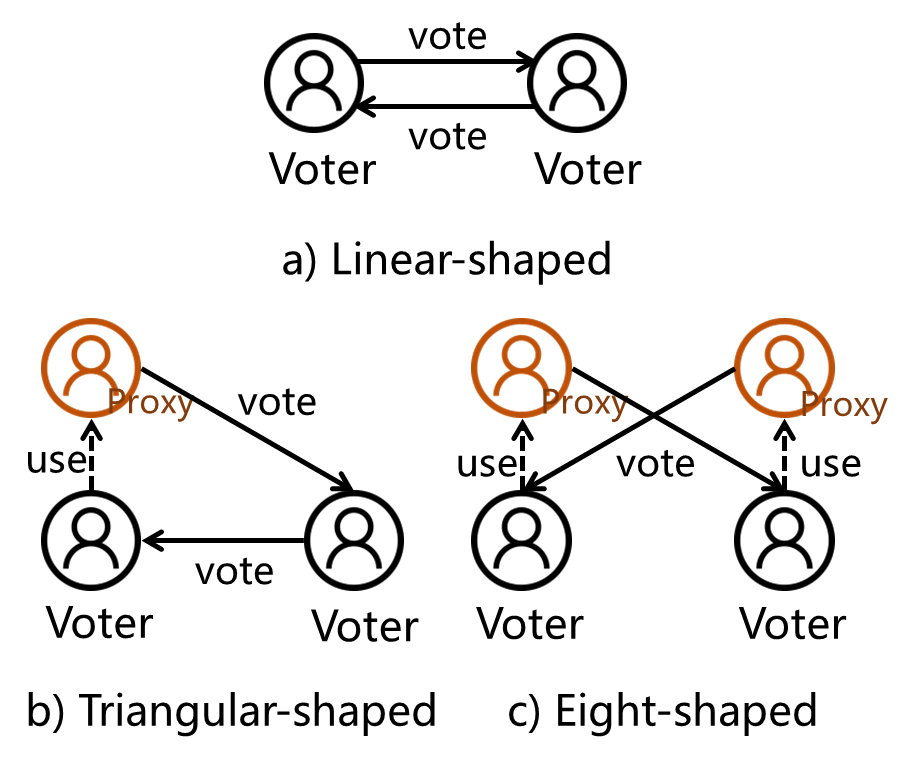}
	}
	\subfigure[Number evolution]{\label{mutual_statistics}
		\includegraphics[scale=0.26]{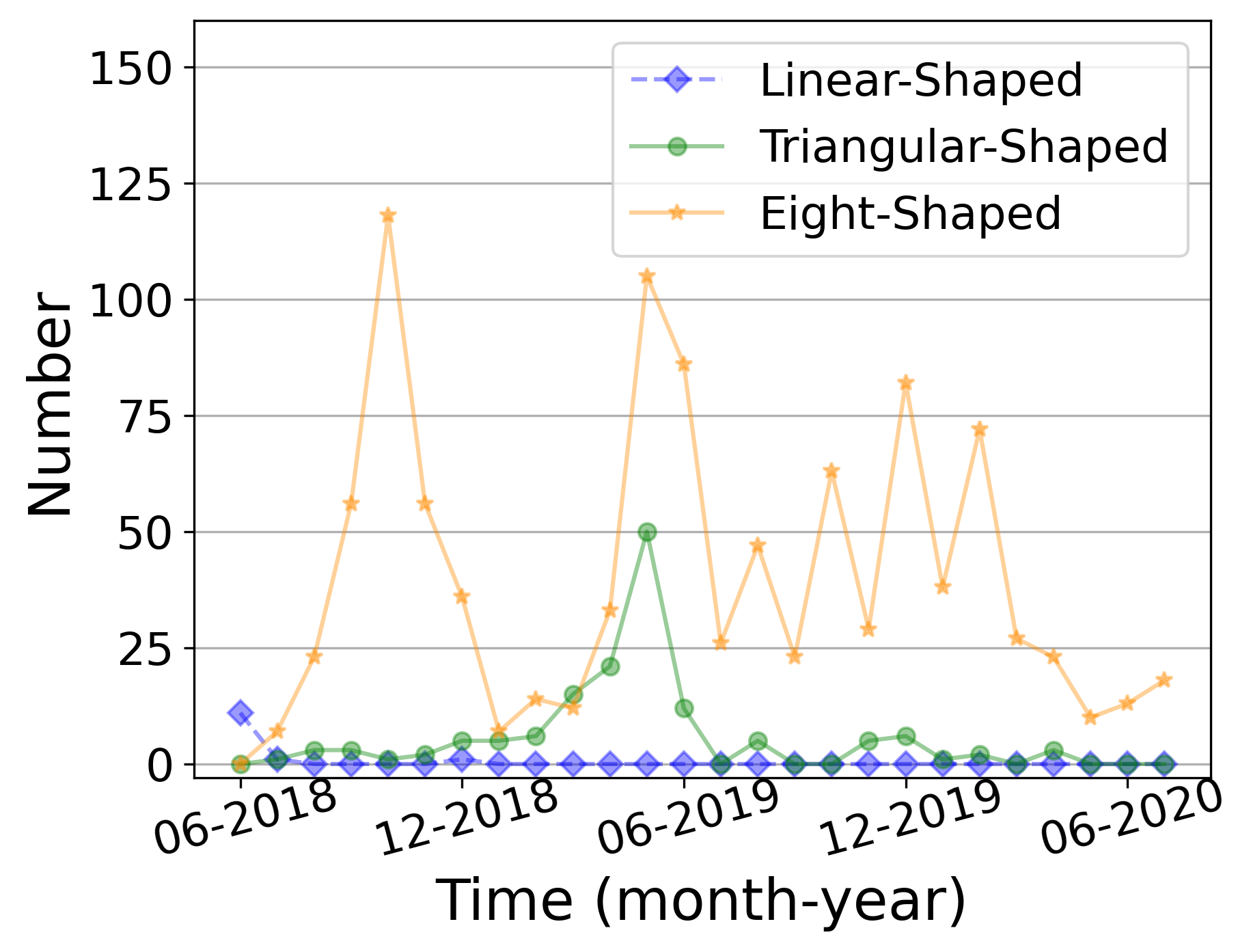}
	}
	\caption{Three mutual voting patterns and their number evolution.}
\end{figure}

In this part, we aim to investigate the mutual voting anomalies, which are suspected of trading votes and sharing the rewards of block production with gang members. We first propose and discuss three mutual voting patterns, and then we develop an algorithm to explore the potential mutual voting anomalies.

\subsubsection{Mutual Voting Pattern Analysis}{\label{DMVNA}}
As shown in Fig. \ref{mutual_voting_pattern}, we consider three types of mutual voting patterns in the analysis: a) \textit{linear-shaped} pattern, in which two voters vote for each other directly to enhance their roles; b) \textit{triangular-shaped} pattern, where a voter $a$ use a proxy to vote for another voter $b$ and receive the vote back from the voter $b$; and c) \textit{eight-shaped} pattern, in which two proxies and two proxied voters are involved and the two proxied voters vote for each other via proxies. Among these three kinds of mutual voting patterns, the linear-shaped pattern and the triangular-shaped pattern are relatively abnormal voting behaviors since there exist votes from non-proxy entities.

The number evolution of these patterns in different months is displayed in Fig. \ref{mutual_statistics}, and the occurring time of the mutual voting actions in each pattern are constrained within seven days. We observe that the most straightforward mutual voting pattern, namely the linear-shaped pattern, occurred mainly between some famous super nodes such as eosnationftw, eoslaomaocom, and argentinaeos in June 2018. For the triangular-shaped pattern, its occurrence number reached a peak in May 2019. By analyzing the voting records, we find that the occurrence of this peak is related to the obvious mutual voting behaviors between games.eos, eos.fish, starteosiobp, and dilidilifans. Firstly, the votes between starteosiobp and eos.fish, games.eos and eos.fish, dilidilifans and eos.fish are in triangular-shaped patterns with proxy start13.io, proxyfordili, and starteos.io respectively. Secondly, dilidilifans and games.eos, starteosiobp and dilidilifans, games.eos and starteosiobp are in eight-shaped patterns via their proxy, which provides further proof for the close relationship among these voters.

As for the eight-shaped pattern, it occurs more frequently than the linear-shaped pattern and triangular-shaped pattern because voting proxies are active. We then investigate the occurrence number peak of the eight-shaped pattern. The maximum peak reached in October 2018, mainly caused by the frequently mutual voting of acroeos12345 and eoseouldotio with a common proxy votetochange. This mutual voting relationship is natural since acroeos12345 and eoseouldotio are both the organizers of the proxy votetochange\footnote{https://www.alohaeos.com/vote/proxy/votetochange}. For the second peak reached in May 2019, we observe that there exists a large connected component formed by the mutual voting relationships of proxied voters. And among the 17 proxied voters in this connected component, the average clustering coefficient is 0.431 and 26 triangles are constructed by their relationships, indicating that these voters tend to gather together.

\subsubsection{Mutual Voting Gang Detection}
To investigate the mutual voting gang anomalies, we build a voting network based on the voting actions, where each node denotes an account and each edge denotes the voting relationship between a pair of accounts. Then, we visualize the voting network by randomly selecting 5,000 edges. As shown in Fig. \ref{fig_voting_network_}, the voting network contains many hub nodes whose one-hop neighborhood (also named egonet) is in a near-star pattern and the one-hop neighbors are almost not connected. While in voting collusion, the neighbors of a node tend to have more connections to maintain their role. Therefore, this kind of voting anomaly can be described by the near-clique pattern (Fig. \ref{fig_near_clique}) that the one-hop neighbors of a node are very well connected. Based on this observation, we first detect the near-clique anomalies among the block producer candidates, and then discover the mutual voting gangs within the voting network of the anomalies. The whole analysis process contains three steps:


\begin{figure}[t]
	\subfigure[The voting network]{\label{fig_voting_network_}
		\includegraphics[scale=0.2]{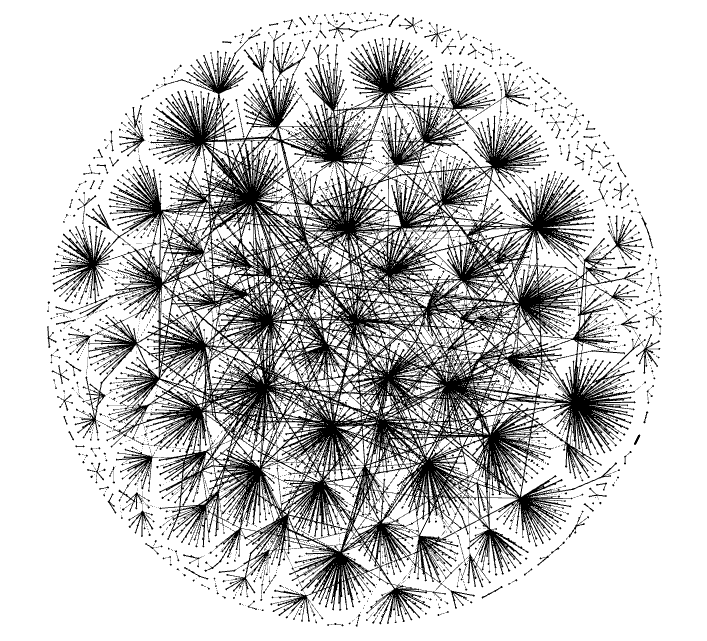}
	}
	\subfigure[An example of near-clique anomaly]{\label{fig_near_clique}
		\includegraphics[scale=0.18, trim=-100 0 -100 20]{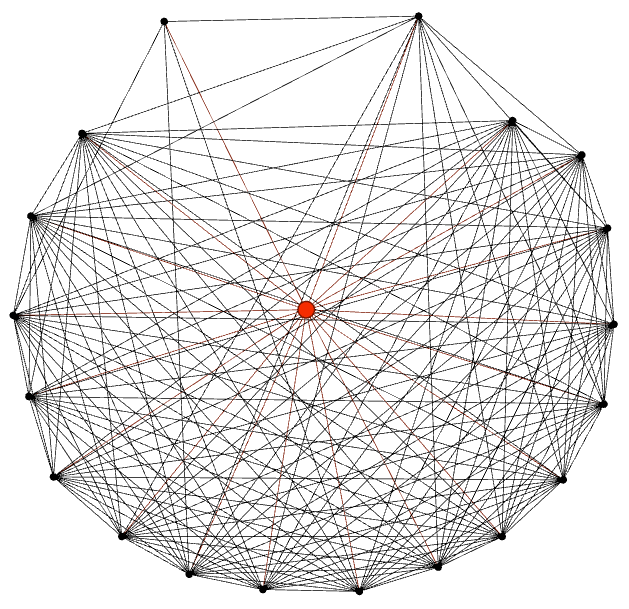}
	}
	\centering
	\caption{Visualization of the voting network (a) and near-clique anomaly (b).}
	\label{fig_voting_network}
\end{figure}

\textbf{Step 1: Near-clique anomaly detection.} Here the OddBall algorithm \cite{akoglu2010oddball} is applied to identify the suspicious block producer candidates which have a relatively closed connected neighborhood. To spot near-clique anomalies, the OddBall algorithm makes use of two features extracted from the egonet of each node, namely $N_i$ (the number of neighbors of node $i$) and $E_i$ (the number of edges in the egonet of node $i$). Based on the observation that $N_i$ and $E_i$ for nodes in a network follow a \textit{Egonet Density Power Law} $E_i\propto N^\alpha_i$ ($1\leqslant\alpha\leqslant 2$) \cite{akoglu2010oddball}, the outliers whose $E_i$ deviates from and is much greater than the expected value $CN^\alpha_i$ can be identified as near-clique anomalies. And the outlierness score of a node can be calculated as follow according to its distance to the fitting line:
\begin{equation}\label{outlierscore}
Score(i)=\frac{max(E_i,CN^\alpha_i)}{min(E_i,CN^\alpha_i)}*log(|E_i-CN^\alpha_i|+1),
\end{equation}
where $\frac{max(E_i,CN^\alpha_i)}{min(E_i,CN^\alpha_i)}$ is the penalty coefficient measures the times that $E_i$ deviates from $CN^\alpha_i$ for a node $i$, and $log(|E_i-CN^\alpha_i|+1)$ is a logarithmic distance measure.

\begin{figure}[t]
	\subfigure[Outlierness score distribution]{\label{fig_outlier_score}
		\includegraphics[scale=0.27, trim= 0 0 0 0]{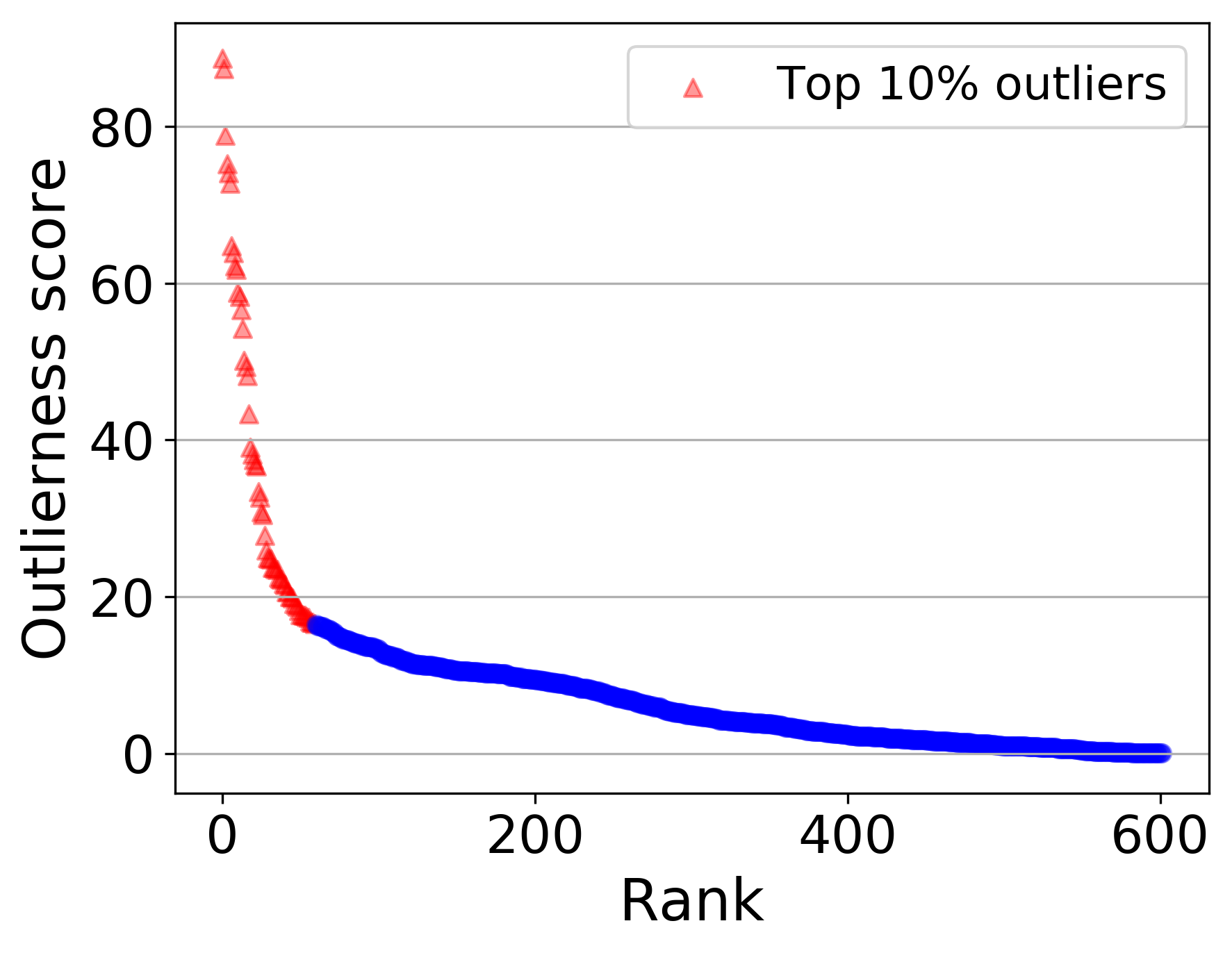}
	}
	\subfigure[The Egonet Density Power Law]{\label{fig_power_law}
		\includegraphics[scale=0.27, trim= 0 0 0 0]{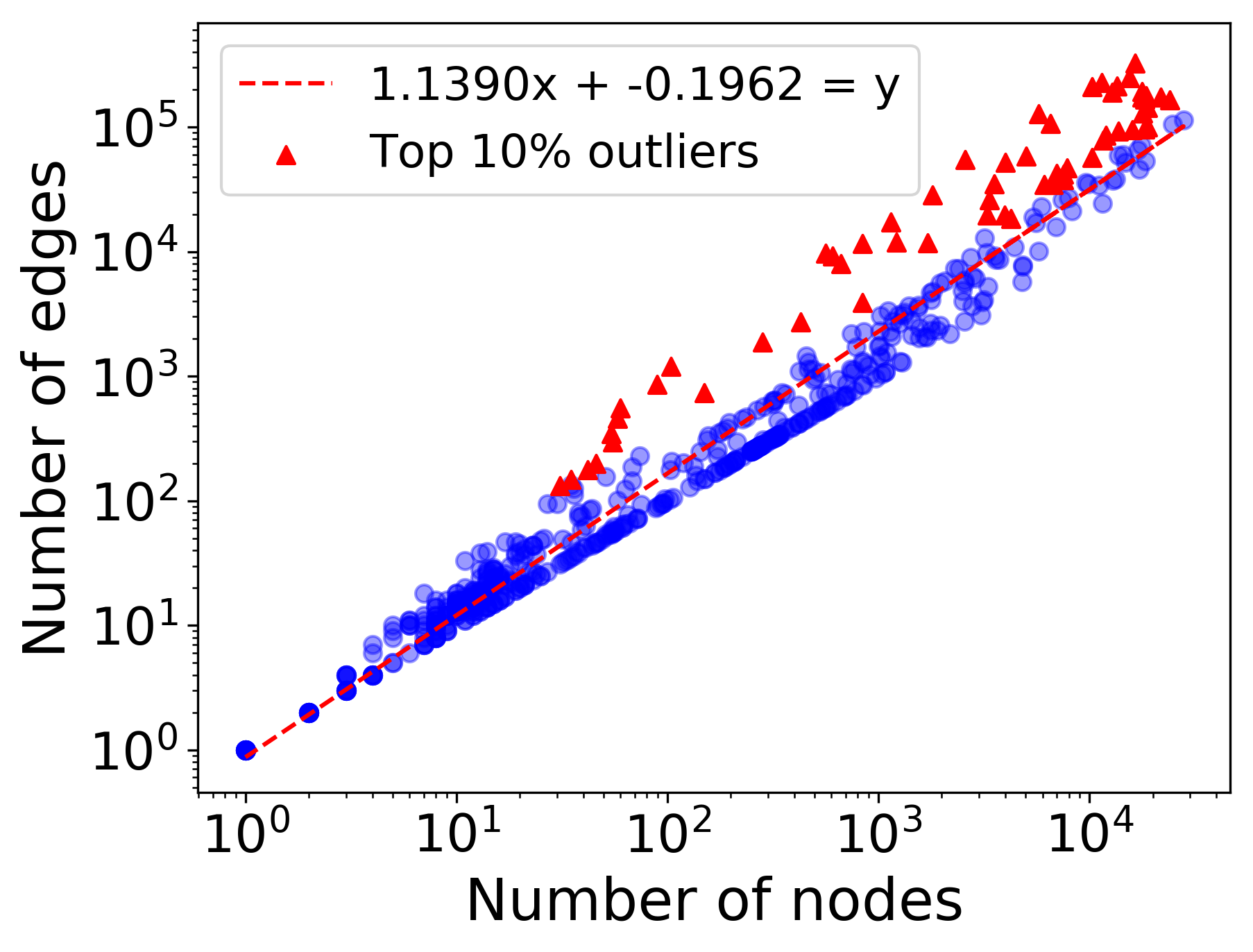}
	}
	\centering
	\caption{Result of the near-clique anomaly detection. (a) shows the distribution of outlierness scores, and (b) shows the visualization of the Egonet Density Power Law, where the red dotted line is the fitting line of power law. The top 10\% outliers in near-clique pattern with the largest outlierness score are assigned red and marked with triangles.}
	\label{fig_OddBall_detection_result}
\end{figure}

\textbf{Step 2: Network reconstruction.} In order to represent the voting intensity between a pair of nodes, we assign a weight to each edge in the network. And the weight $w_{ij}$ of the edge between nodes $i$ and $j$ is decided by the voting intensity from $i$ to $j$ (denoted by $I_{ij}$) and the voting intensity from $j$ to $i$ (denoted by $I_{ji}$), i.e., $w_{ij} = I_{ij} + I_{ji}$.
Here the voting intensity is calculated according to the voting frequency, duration, and voting power as Equation (\ref{weight2}).
\begin{equation}\label{weight2}
I_{ij} = \frac{1}{3} * (\frac{F_{ij}}{\sum_{u\in N(i)}F_{iu}} + \frac{T_{ij}}{\sum_{u\in N(j)}T_{uj}} + \frac{P_{ij}}{\sum_{u\in N(j)}P_{uj}}),
\end{equation}
where $N(\cdot)$ represents the neighbors of a node, $F_{ij}$ represents the voting frequency from $i$ to $j$, $T_{ij}$ is the cumulative voting time duration and $P_{ij}$ is the average voting weight from $i$ to $j$. The item about the voting frequency measures the willingness of $i$ voting to $j$, and the items about the voting time duration and average voting weight measure $i$'s voting share of $j$. Then, we reconstruct the voting network by only keeping the relationships between the block producer candidates in the egonet of the detected accounts with near-clique neighborhood, paving the way for investigating the mutual voting gangs.

\textbf{Step 3: Community detection.} We apply the Louvain algorithm \cite{Blondel_2008} on the weighted network to detect suspicious voting gangs, and the community detection results are obtained according to the voting intensity calculated by Equation (\ref{weight2}) and modularity optimization. After obtaining the clustering results, we weed out the accounts with only one edge in the reconstructed network since these accounts have less possibility to participate in voting collusion in their cluster. And the outputs are clusters with more than one node.

\textit{Results:} The distribution of the calculated outlinerness score is shown in Fig. \ref{fig_outlier_score}, and here we choose the top 10\% outliers with the highest outlierness score as the detected anomalies (totally 60 accounts), which are assigned red and marked with triangle in the figure. Fig. \ref{fig_power_law} shows the visualization of the \textit{Egonet Density Power Law}. We can observe that these detected anomalies deviate from and are above the fitting line of the \textit{Egonet Density Power Law}. The reconstructed network based on the detected anomalies contains 334 nodes and 1,604 undirected edges. And finally, we obtain 11 abnormal voting gangs by community detection.

\begin{figure}[t]
	\centerline{\includegraphics[scale=0.16]{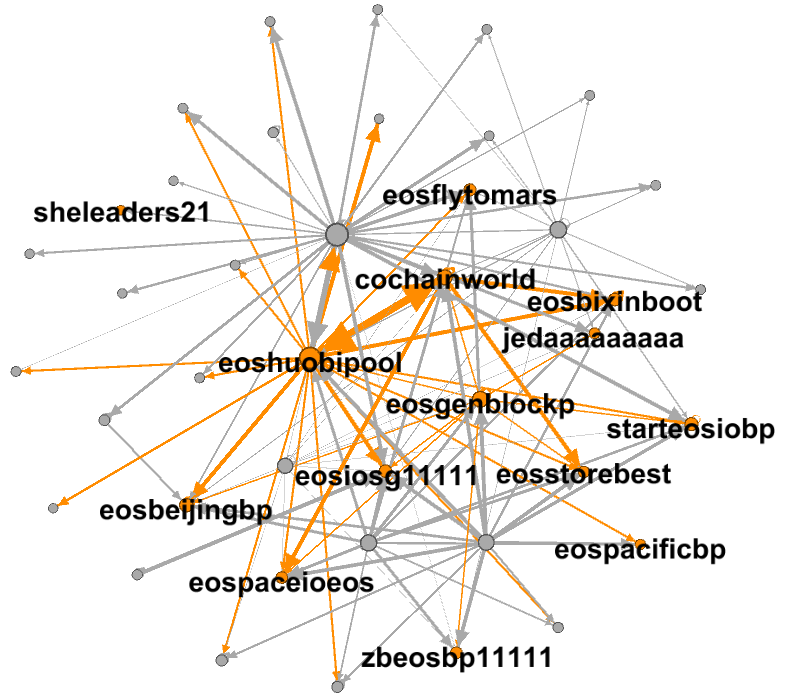}}
	\caption{Visualization of the voting relationships in the detected cluster of \textit{`eoshuobipool'}, where the accounts reported in \cite{Huobidataleak} are labeled and assigned the color orange.}
	\label{fig_huobicluster}
\end{figure}

In particular, we find out that a detected cluster, which includes 41 accounts, has 14 accounts coinciding with more than half of the accounts allegedly involved in mutual voting reported in \cite{Huobidataleak}. The voting relationships in this cluster are shown in Fig. \ref{fig_huobicluster}, where the detected labeled accounts and their relationships are assigned the color orange. As we can see in this figure, though there are some accounts having not been reported in \cite{Huobidataleak}, they have strong voting relationships with the cluster members. Moreover, 24 of the 26 reported accounts have been detected in the output clusters. Besides, we observed that there is a huge detected cluster consisting of 105 accounts. Among these accounts, 102 are named in a regular pattern with `hayd' or `g44t' as the first four letters and `ge' as the last two letters, such as \textit{`haydiojxgege'}, \textit{`haydiojygage'}, \textit{`g44tomrygige'} and \textit{`g44tomzugqge'}. And 2 of the last 3 accounts are created by two accounts in this naming rule. These observations illustrate that this three-step method can help us notice some abnormal mutual voting gangs.

\section{Discussion} \label{implications}
Previous sections have presented a decentralization evolution analysis on EOSIO, a typical DPoS-based blockchain system, and reported some abnormal voting phenomena in the system. In this section, we discuss some implications from our findings for decentralization enhancement in DPoS-based blockchain systems.

\textbf{First}, voters in EOSIO only account for a small fraction of stakeholders, and their voting power is distributed unevenly, which means a few voters possessing a large number of stakes are sufficient to centralize the network.
Moreover, since each voter can vote for up to 30 block producer candidates at the same time with no discount on the voting weight that each candidate can receive, the cost of voting manipulation for malicious account gangs is relatively low. Therefore, except to encourage the stakeholders to be active in the governance of EOSIO, the community can also look for a more reasonable value for the number of votes allowed per account. Existing work \cite{jeong2020centralized} has studied how to find the optimal number of votes allowed per account with governance game. Besides, another possible solution can be the introduction of an attenuation coefficient when a user votes for multiple candidates. In this way, a voter can hold different ratings for the multiple chosen candidates and give corresponding voting weights based on the ratings.

\textbf{Second}, voting proxies are popularly used recently, making the votes more centralized. Some voting proxies even attract new users by claiming that delegating the voting power to proxies can bring new users high profit with zero risk. However, these voting proxies are likely to become large staking pools, which not only violates the original design intention of voting proxies but also raises some security concerns similar to those brought by mining pools in Bitcoin \cite{9160462}, such as 51\% attacks. With the proposed clustering method, we have identified some suspicious voting proxy gangs in which the members have similar voting behaviors in both time and space. The proxies in these detected clusters probably have made use of the collected voting power to vote for their alliance members. We suggest that users can utilize our method to investigate the voting behaviors of voting proxies before choosing a voting proxy. 

\textbf{Third}, from the collected data, we observe that the block producer set has small variations, and this fact makes the system rather vulnerable to centralized control and malicious attacks. For example, Xu et al. \cite{xu2018eos} proposed a type of voting attack executed by block producers\textemdash the block producers can collude to blacklist the voters with large stakes which will threaten the authority of the block producers themselves. 
Since there is no protocol to prevent this type of vulnerability, it is vital to select reliable nodes in block producer election. As introduced earlier, our method can detect the block producers participating in mutual voting and thus can provide a reference for analyzing block producers in DPoS-based blockchain.

\textbf{Last but not least}, our analysis methods can be generalized to other DPoS-based blockchain systems like Tron, Binance Smart Chain and OKExChain \cite{OKC2020} since these systems employ similar settings in their voting mechanism.


\section{Related Work}\label{RW}
We present the related work in this section, which mainly includes the decentralization analyses of DPoS-based blockchain systems and existing analyses in EOSIO. 
\subsection{Decentralization Analysis of DPoS-based Blockchains}
Decentralization is an important characteristic that can tell blockchain systems from traditional centralized systems. At present, there have been a few studies on analyzing the decentralization of blockchain systems. Most of the existing researches pointed out that Bitcoin and Ethereum have not achieved true decentralization in terms of mining power, network resources such as bandwidth, etc \cite{7176229,10.1007/978-3-662-58387-6_24,9160462}. Since PoW-based blockchains become more and more centralized, Chen et al. \cite{chen2020endex} proposed tools to timely quantify the decentralization in terms of mining power in PoW-based blockchain systems.
While for DPoS-based blockchains, the block production process does not require every node in the network to participate in block production directly. Instead, users can vote for a specified number of block production candidates, and the elected block producers generate blocks in turns. Thus in DPoS-based blockchains, the mining power is a weak influence factor of decentralization.

For DPoS-based blockchain systems, Jeongy \cite{jeong2020centralized} explored the relationship between decentralization and the number of votes allowed per account, and found that the ``one vote per account'' rule for centralization mitigation may be not necessary. 
Rebello et al. \cite{rebellosecurity} analyzed several quorum-based consensus protocols including the EOSIO protocol, and pointed out some clear vulnerabilities of these protocols.
Kwon et al. \cite{10.1145/3318041.3355463} measured the degree of decentralization in several blockchain systems based on PoW, PoS, and DPoS with the entropy metric and explain why it is difficult to design a fully decentralized system. 
Li and Palanisamy \cite{10.1007/978-3-030-59638-5_2} conducted a comparison of decentralization between Bitcoin and Steem \cite{10.1145/3292522.3326041}, which are based on PoW and DPoS respectively. They found that compared with Steem, Bitcoin tends to be less decentralized in general and more decentralized among top stakeholders via some distributions and the entropy metric.
However, there is a lack of in-depth analyses studying the decentralization evolution and uncovering the abnormal voting phenomena in DPoS-based systems.

\subsection{EOSIO Analysis}
Nowadays, blockchain has attracted intensive interests of researchers. Existing studies cover several aspects including blockchain architecture and performance \cite{zheng2017overview,9031420}, privacy and security \cite{khalilov2018survey,Jieli9332279}, smart contracts and transactions \cite{10.1145/3377811.3380364,Wu2020AnalysisOC}, etc. 
As a successful and typical DPoS-based blockchain system, EOSIO has received attention of scholars these years.

In terms of architecture and security, Xu et al. \cite{xu2018eos} provided a comprehensive analysis on EOSIO's architecture, performance, and economy.
Lee et al. \cite{238604} introduced four attacks in EOSIO, whose root cause lies in the characteristics of EOSIO.  
Quan et al. \cite{quan2019evulhunter} proposed a static analysis tool to detect the fake-transfer vulnerabilities of EOSIO's smart contracts.
He et al. \cite{he2020security} proposed a tool that can detect four popular vulnerabilities in EOSIO smart contracts at the WebAssembly code level.
Lee et al. \cite{lee2019push} proposed a kind of attack that can disturb the fairness of incentive policy by manipulating the block production schedule in EOSIO, and this attack can bring additional rewards or losses to block producers.
In terms of blockchain transaction analysis, Huang et al. \cite{10.1145/3392155} investigated EOSIO and the associated DApps via measurement study. With a focus on security issues, they discovered some real-world attacks and developed detection techniques for fraudulent activities in EOSIO.
Zheng et al. performed \cite{zheng2020xblock} a statistical analysis on 7 well-processed datasets in EOSIO, and outlined some possible research directions based on the proposed datasets. 
Zhao et al. \cite{10.1007/978-981-15-9213-3_37} abstracted the records of four major activities in EOSIO as networks, and obtained some interesting observations via network metric analysis. 
However, none of them quantify the decentralization characteristic and investigate the abnormal voting phenomena based on the real transaction data in EOSIO.

\section{Conclusion and Future Work}\label{Conclusion}
This paper performed the first decentralization evolution study on DPoS-based blockchains. Based on the blockchain data from EOSIO, a typical DPoS-based blockchain system, we characterized the activities of voters, voting proxies, and block producers participating in the DPOS consensus process and obtained many insightful findings. Moreover, we presented methods to discover the abnormal voting gangs with similar voting behaviors and mutual voting behaviors. Some suspected voting manipulation phenomena have been revealed in our analysis. Besides, we provided some implications for enhancing the decentralization of EOSIO, which can also provide a reference for other DPoS-based blockchain systems. 

For future work, we will expand our study in two directions. First, there may exist other complex abnormal voting patterns needing to be explored, we plan to investigate more abnormal voting phenomena in the system by combining with the token trading data and the resource trading data in EOSIO. Second, we want to extend this analysis to other DPoS-based blockchain systems, and build a decentralization monitoring platform to timely quantify the degree of decentralization and capture the abnormal phenomena in different DPoS-based blockchain systems.


\bibliography{ref}
\bibliographystyle{IEEEtran}


%

\begin{IEEEbiography}
	[{\includegraphics[width=1in,height=1.25in,clip,keepaspectratio]{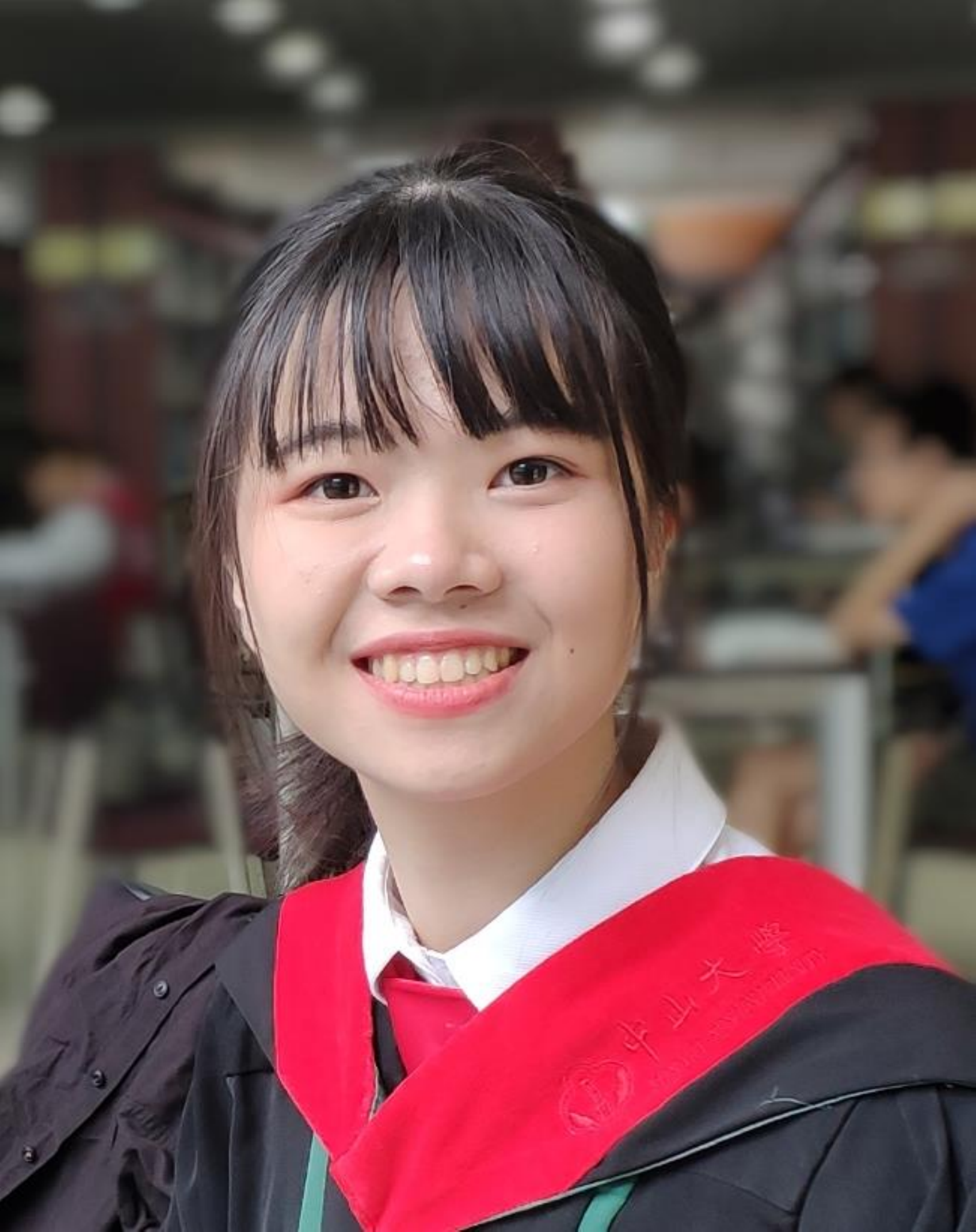}}]{Jieli Liu} received her B.Eng. in Software Engineering from Sun Yat-sen University, Guangzhou, China, in 2019. She is currently studying toward a Ph.D. degree in the School of Software Engineering, Sun Yat-sen University. Her current research interests include blockchain, network science, data mining, and machine learning with graphs.
\end{IEEEbiography}

\begin{IEEEbiography}
	[{\includegraphics[width=1in,height=1.25in,clip,keepaspectratio]{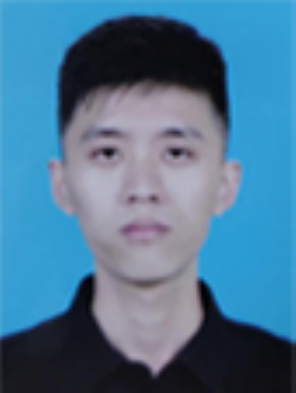}}]{Weilin Zheng} is currently pursuing the M.Eng. degree with the School of Computer Science and Engineering, Sun Yat-sen University, Guangzhou, China. His research interests include performance monitoring and optimization, blockchain computing power utilization, blockchain data analysis, and blockchain-based decentralized applications.
\end{IEEEbiography}

\vspace{-4ex}
\begin{IEEEbiography}
	[{\includegraphics[width=1in,height=1.25in,clip,keepaspectratio]{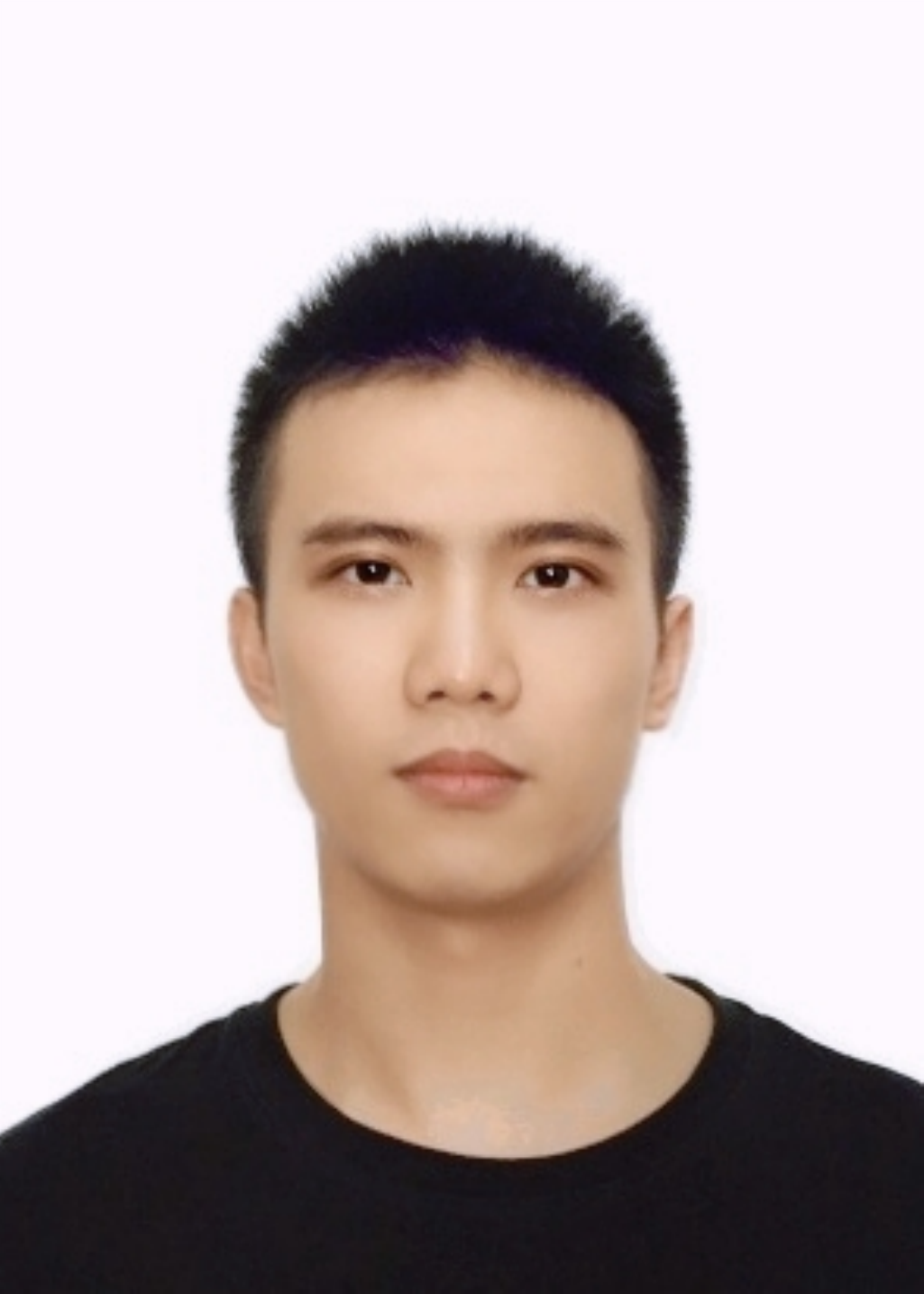}}]{Dingyuan Lu} received the B.Eng. in computer science and technology from Shanxi University, Taiyuan, China, in 2018. He is currently pursuing an M.Eng. degree in the School of Computer Science and Engineering, Sun Yat-sen University, Guangzhou, China. His current research interests include applications of network science, blockchain, and machine learning with graphs.
\end{IEEEbiography}

\vspace{-4ex}
\begin{IEEEbiography}
	[{\includegraphics[width=1in,height=1.25in,clip,keepaspectratio]{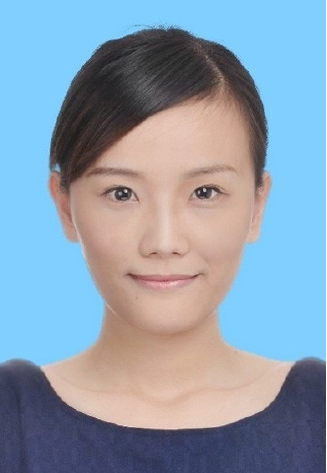}}]{Jiajing Wu} (S'11--M'14--SM'19) received the B.Eng. degree in communication engineering from Beijing Jiaotong University, Beijing, China, in 2010, and the Ph.D. degree from Hong Kong Polytechnic University, Hong Kong, in 2014. She was awarded the Hong Kong Ph.D. Fellowship Scheme during her Ph.D. study in Hong Kong (2010--2014). 
	
	In 2015, she joined Sun Yat-sen University, Guangzhou, China, where she is currently an Associate Professor. Her research focus includes blockchain, graph mining, network science. She serves as an Associate Editor for {\sc IEEE Transactions on Circuits and Systems II: Express Briefs.}
\end{IEEEbiography}

\vspace*{-4ex}
\begin{IEEEbiography}
	[{\includegraphics[width=1in,height=1.25in,clip,keepaspectratio]{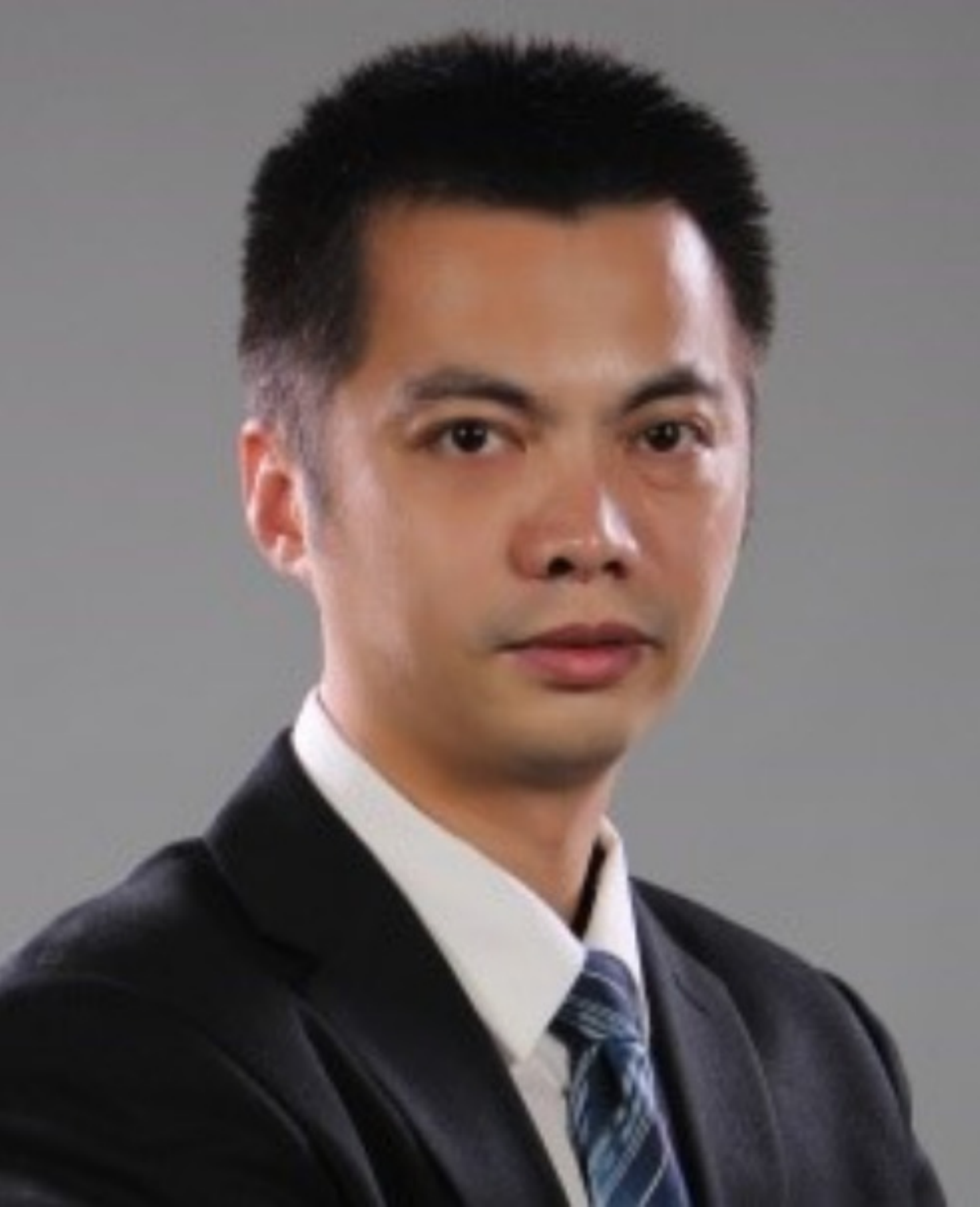}}]{Zibin Zheng} received a Ph.D. degree from the Chinese University of Hong Kong in 2011. 
	
	He is currently a Professor of Data and Computer Science with Sun Yat-sen University, China. He serves as the Chair of the Software Engineering Department, Pearl River Young Scholars, and the Founding Chair of the Services Society Young Scientists Forum (SSYSF). In the past five years, he published over 120 international journal and conference papers, including three ESI highly-cited papers, 40 {\sc ACM/IEEE Transactions} papers. According to Google Scholar, his papers have more than 6300 citations, with an H-index of 41. His research interests include blockchain, services computing, software engineering, and financial big data. He was the recipient of several awards, including the outstanding Thesis Award of CUHK, in 2012, the ACMSIGSOFT Distinguished Paper Award at ICSE2010, the Best Student Paper Award at ICWS2010, and IBM Ph.D. Fellowship Award. He served as CollaborateCom’16 General co-Chair, ICIOT’18 PC co-Chair, and IoV’14 PC co-Chair.
\end{IEEEbiography}
%
%
%
%

\vfill

\end{document}